\RequirePackage{fix-cm}
\documentclass[twocolumn]{svjour3}          
\smartqed  

\usepackage{graphicx}
\usepackage{xcolor}
\usepackage{amsmath}
\usepackage{textcomp}
\usepackage{cite}

\usepackage[hyphens]{url}
\usepackage{mathbbol}
\usepackage{hyperref}

\newcommand{\done}[1]{} 

\journalname{Computing and Software for Big Science}
\begin{document}

\title{Getting High:
High Fidelity Simulation of High Granularity Calorimeters with High Speed
}


\author{Erik Buhmann \and Sascha Diefenbacher \and Engin Eren \and Frank Gaede \and Gregor Kasieczka \and Anatolii Korol \and Katja Kr\"uger}


\institute{
%
E. Buhmann, S. Diefenbacher, and G. Kasieczka \at
              Institut f\"ur Experimentalphysik, Universit\"at Hamburg, Germany
              \email{sascha.daniel.diefenbacher@uni-hamburg.de}           
           \and
           E. Eren, F. Gaede, and K. Kr\"uger \at
           Deutsches Elektronen-Synchrotron, Germany\\
           \email{engin.eren@desy.de}
           \and
           A. Korol \at
           Taras Shevchenko National University of Kyiv, Ukraine
}

\date{Received: date / Accepted: date}





\maketitle
\begin{abstract}

Accurate simulation of physical processes is 
crucial for the success of modern particle physics.
However, simulating the development and interaction of particle showers with calorimeter detectors is a time consuming process and  drives the computing needs of large experiments at the LHC and future colliders. Recently, generative machine 
learning models based on deep neural networks have shown promise in speeding up this task by several orders of magnitude. We investigate 
the use of a new architecture --- the Bounded Information Bottleneck Autoencoder ---
for modelling electromagnetic showers in the central region of the Silicon-Tungsten calorimeter
of the proposed International Large Detector. Combined with a novel second post-processing network, this approach
achieves an accurate simulation of differential distributions 
including for the first time  the shape of the minimum-ionizing-particle peak
compared to a full Geant4 simulation for a high-granularity calorimeter with
27k simulated channels. The results are validated by comparing to established architectures. 
Our results further strengthen the case of using generative networks for fast simulation and demonstrate that 
physically relevant differential distributions can be described with high accuracy.



\end{abstract}
\newcommand{\sd}[1]{{\color{blue}(SD: #1)}}

\section{Introduction}
\label{sec:intro}

Precisely measuring nature’s fundamental parameters and discovering new 
elementary particles in modern high energy physics is only made possible
by our deep mathematical understanding of the Standard Model
and our ability to reliably simulate interactions of these particles 
with complex detectors. While essential for our scientific progress,
the production of these simulations is increasingly costly.
This cost is already a potential bottleneck at the LHC,
and the problem will be exacerbated by higher luminosity,
larger amounts of pile-up and more complex and granular detectors at the High-Luminosity
LHC and planned future colliders. 
A promising way to accelerate the simulation is offered by generative machine learning models and was pioneered in Ref.~\cite{CaloGAN2}. 
The present work focuses on simulating a very high-resolution calorimeter prototype with greater fidelity of
physically relevant distributions, paving the road for practical applications~\footnote{Implementations of
the network architectures as well as instructions to produce training data are available on \\
\url{https://github.com/FLC-QU-hep/getting_high}.}.


Advanced machine learning methods, based on deep neural networks, are rapidly transforming and improving the way to
explore the fundamental interactions of nature in particle physics --- see for example Ref.~\cite{jets_comparison} 
for a recent overview of neural network architectures developed to identify hadronically decaying top quarks.
However, we are only beginning to explore the potential benefits from unsupervised techniques designed to  
model the underlying high-dimensional density distribution of data. This allows, e.g., anomaly detection
algorithms to identify signals from new physics theories without making  specific model assumptions~\cite{weak6,weak7,weak8,koalahunting,TaoNovelty,tagntrain,anode,SALAD,ALAD,ATLASAnomaly}.
Furthermore, once the phase space density is encoded in a neural network, it can be sampled from very efficiently. 
This makes synthetic models of particle interactions  many orders of magnitude faster than
classical approaches, where for example for a particle showering in a calorimeter many secondary shower
particles have to be created and individually tracked through the material of the detector according to the underlying physics processes.

Calorimeters are a crucial part of experiments in high energy physics, where the incident primary particles create showers of secondary
particles in dense materials that are used to measure the energy. In sandwich calorimeters, layers of dense materials are interleaved with
sensitive layers recording energy depositions from secondary shower particles mostly from ionization. The details of the shower development via
creation of secondary particles as well as their energy loss is typically simulated with great accuracy using the Geant4~\cite{g4} toolkit.

The crucial role of calorimeter simulation as a time-consuming bottleneck in the  simulation chain at the LHC is well established.
For example, the ATLAS experiment uses more than half of its total CPU time on the LHC Computing Grid for Monte Carlo simulation, which in turn is entirely
dominated by the calorimeter simulation~\cite{atlas_fastsim}. 



While generative neural network techniques promise enormous speed-ups for simulating the calorimeter response, it is
of extreme importance that all relevant physical shower properties are reproduced accurately in great detail.
This is particularly challenging for highly granular calorimeters, with a much higher spatial resolution,
foreseen for most future colliders. Such concepts, as developed for the International Linear Collider (ILC), are also being used to upgrade
detectors at the LHC for upcoming data-taking periods. One prominent example is the calorimeter endcap upgrade of the CMS experiment~\cite{hgcal_tdr} with about 6 million readout channels.
These factors make the timely development of precise simulation tools for high-resolution 
detectors relevant and motivate our investigation of a prototype calorimeter for the International Large Detector (ILD).

Outside of particle physics, generative adversarial neural networks~\cite{GAN} (GANs) have been used to produce synthetic data 
--- such as photo-realistic images~\cite{StyleGAN} --- with great success. A traditional GAN consists of two networks, a generator and a discriminator separating artificial samples from real ones, which are trained against each other. An alternative to GANs for simulation are Variational Autoencoders~\cite{VAE} (VAE). 
A VAE consists of an encoder mapping from input data to a latent space, and a decoder, which maps 
from the latent space to data. 
If the probability distribution in latent space is known, it can be sampled from and
used to generate synthetic data. 
A third path towards generative models is offered by normalizing flows~\cite{NICE,RealNVP,rezende2015variational,papamakarios2019normalizing,KyleFlow}. In such models, a simple base probability distribution is transformed by a series of invertible mappings into  a complex shape. 

Recently, a novel architecture unifying several generative models such as GANs, VAEs, and others
was proposed: the Bounded-Information-Bottleneck auto\-en\-co\-der (BIB-AE)~\cite{BIB-AE}. We will show that by using a modified BIB-AE for generation we
can accurately model all tested relevant physics distributions to a higher degree than achieved by traditional GANs. A detailed introduction to this 
architecture is provided in Section~\ref{sec:BIBAE}.

Specifically in particle physics, first results for the simulation of calorimeters focused on GANs achieved an impressive speed-up by up to five orders of magnitude compared
to Geant4~\cite{CaloGAN2,CaloGAN,CaloGAN3}.
Similarly, an approach using a Wasserstein-GAN (WGAN) architecture achieved realistic modeling of particle showers in
air-shower detectors~\cite{ErdmannWGAN} and a high granularity sampling calorimeter~\cite{ErdmannWGAN2}.
In the context of future colliders, an architecture inspired by GANs was used 
for the fast simulation of showers in a high granularity electromagnetic calorimeter~\cite{Sofia}.
Generative models based on VAE and WGAN architectures were studied for concrete application by the ATLAS collaboration~\cite{ATLAS_Gen,ATLAS_Gen2, ATLAS_Gen3}.

Beyond producing calorimeter showers, generative models in HEP have also been 
explored for modeling
muon interactions with a dense target~\cite{SHiPGAN},
parton showers~\cite{Bothmann,Monk,JUNIPR,LundCycleGAN},
phase space integration~\cite{badger,Klimek:2018mza,Bendavid:2017zhk,Bothmann:2020ywa},
event generation~\cite{PandolfiGAN,Otten:2019hhl,KaustuvGAN,DiJetGAN,LHCGAN,FlowGeneration},
event subtraction~\cite{ButterSubtraction} and unfolding~\cite{GANFolding}.

The rest of this paper is organised as follows: in Section~\ref{data} we introduce the concrete
problem and training data, in Section~\ref{models} the used generative architectures are discussed,
and in Section~\ref{results} the obtained results are presented and compared. Finally, Section~\ref{closing}
provides conclusions and outlook.




\section{Data Set}
\label{data}

\begin{figure}[b]
    \centering
    \includegraphics[width=0.42\textwidth]{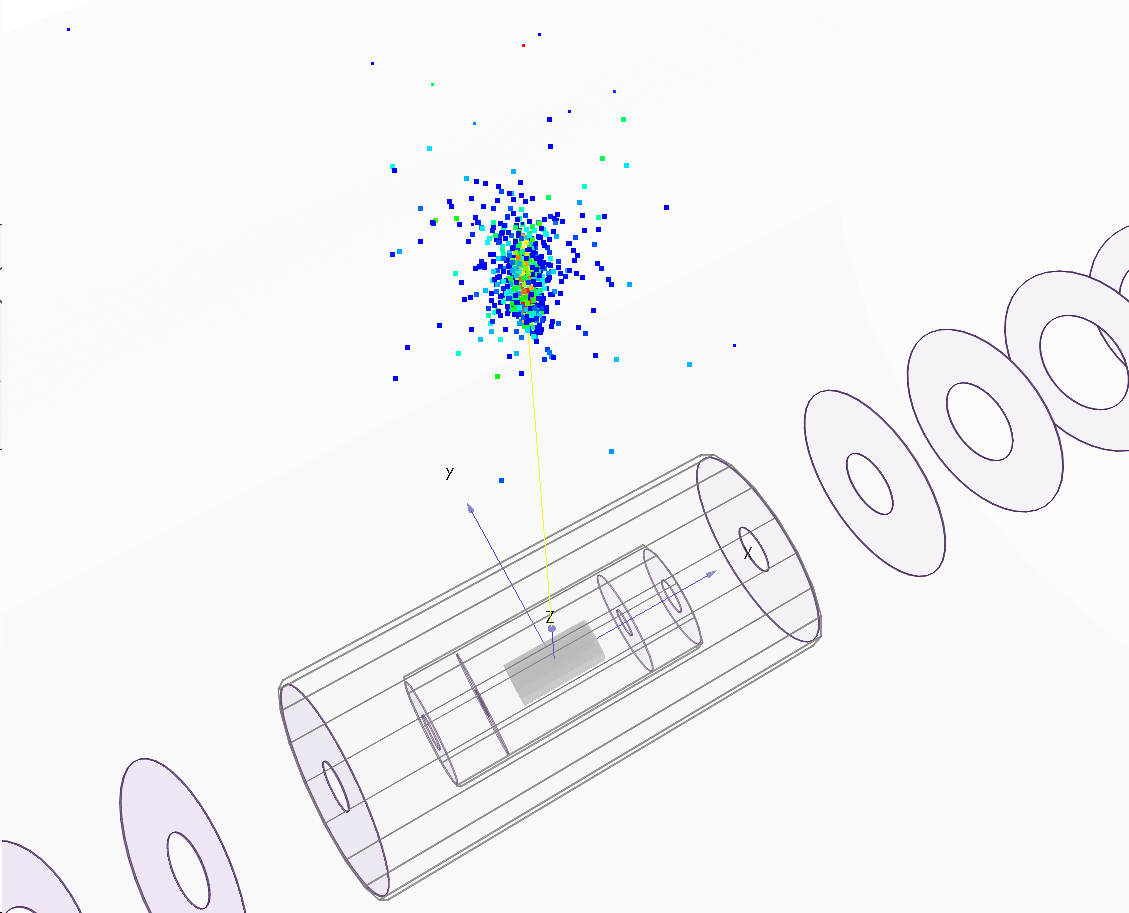}
    \caption{A simulated 60 GeV photon shower in the ILD detector, as used in the training data.
    }
    \label{fig:ecal}
\end{figure}

The ILD~\cite{ILD-IDR} detector is one of two detector concepts proposed for the ILC. It is optimized for
Particle Flow, an algorithm that aims at reconstructing every individual particle in order to optimize the overall detector resolution. ILD combines high-precision tracking and vertexing capabilities with very good hermiticity and highly granular
electromagnetic and hadronic calorimeters. For this study, one of the two proposed electromagnetic calorimeters for ILD, the Si-W ECal is chosen.
It consists of 30  active silicon layers in a tungsten absorber stack with 20 layers of $2.1~\textrm{mm}$ followed by 10 layers of  $4.2~\textrm{mm}$
thickness respectively. The silicon sensors have $5\times 5~\textrm{mm}^2$ cell sizes. 
Throughout this work, we project the sensors onto a rectangular grid of  $30\times30\times30$~cells. Each cell
in this grid corresponds to exactly one sensor. As the underlying geometry of sensors in a
realistic calorimeter prototype is not exactly regular, we will encounter some effects of this staggering. 
This makes the learning task more challenging for the network, but does not pose a fundamental problem.
Architectures that more accurately encode irregular calorimeter geometries in neural networks exist~\cite{CaloGraphs}, but are not the focus of this work.

\begin{figure}[b]
    \centering
    \includegraphics[width=0.35\textwidth]{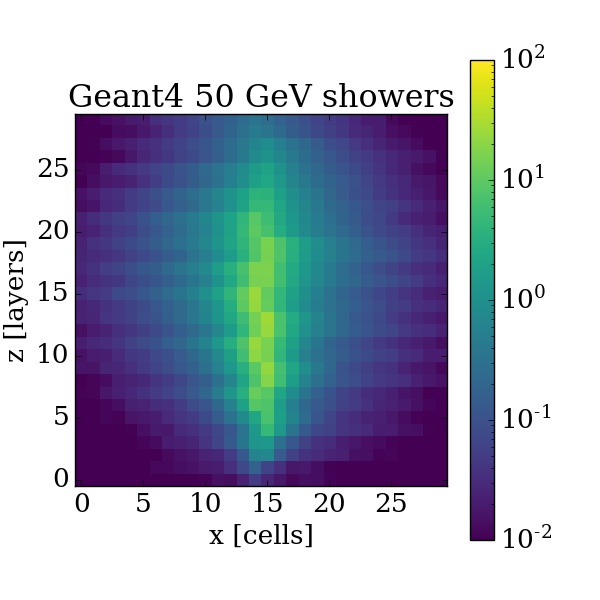}
    \caption{
    Overlay of 2000 projections of 50~GeV Geant4 photon showers along the y direction.}
    \label{fig:showers_projection}
\end{figure}

ILD uses the iLCSoft~\cite{ilcsoft} ecosystem for detector simulation,
reconstruction and analysis. For the full simulation with Geant4, a detailed and realistic detector model implemented in DD4hep~\cite{dd4hep} is used.
The training data of photon showers in the ILD ECal are simulated with Geant4 version 10.4 (with \textrm{QGSP\_\,BERT} physics list) and DD4hep version 1.11.
The photons are shot at perpendicular incident angle into the ECal barrel with energies uniformly\footnote{Due to technical issues with the Geant4 generation step, the produced sample has a difference in statistics of 1$\%$ between the lowest and highest energies} distributed between 10-100 GeV. All incident photons are aimed at the $x-y$ center of the grid --- i.e. at
the point in the middle between the four most central cells of the front layer.
An example event display showing such a photon shower is depicted in Figure~\ref{fig:ecal}. 
%




The incoming photon enters from the bottom at $z=0$ and traverses along the z-axis, hitting cells in the center of the $x-y$ plane.
No variations of the incident angle and impact point are performed in this study. 
The overlay of 2000 showers summed over the y-axis
is shown in Figure~\ref{fig:showers_projection}. As can be seen, 
the cells in the ILD ECal are staggered due to the specific barrel geometry.
The whole data set for training consists of 950k showers with continuous energies between 10-100 GeV. 
For the evaluations we generated additional, statistically independent, sets of events: 40k events uniformly distributed between 10-100 GeV and 
4k events each at discrete energies in steps of 10~GeV between 20 and 90~GeV.


\section{Generative Models}

\label{models}

Generative models are designed to learn an underlying data distribution in a way that
allows later sampling and thereby producing new examples. In the following, we first
present two approaches --- GAN and WGAN --- which 
represent the state-of-the-art in generating calorimeter data and which we use to benchmark our
results. We then introduce BIB-AE as a novel approach to this problem and discuss further 
refinement methods to improve the quality of generated data.




\subsection{Generative Adversarial Network}

\label{sec:GAN}

\begin{figure}[t]
    \centering
    \includegraphics[width=0.48\textwidth]{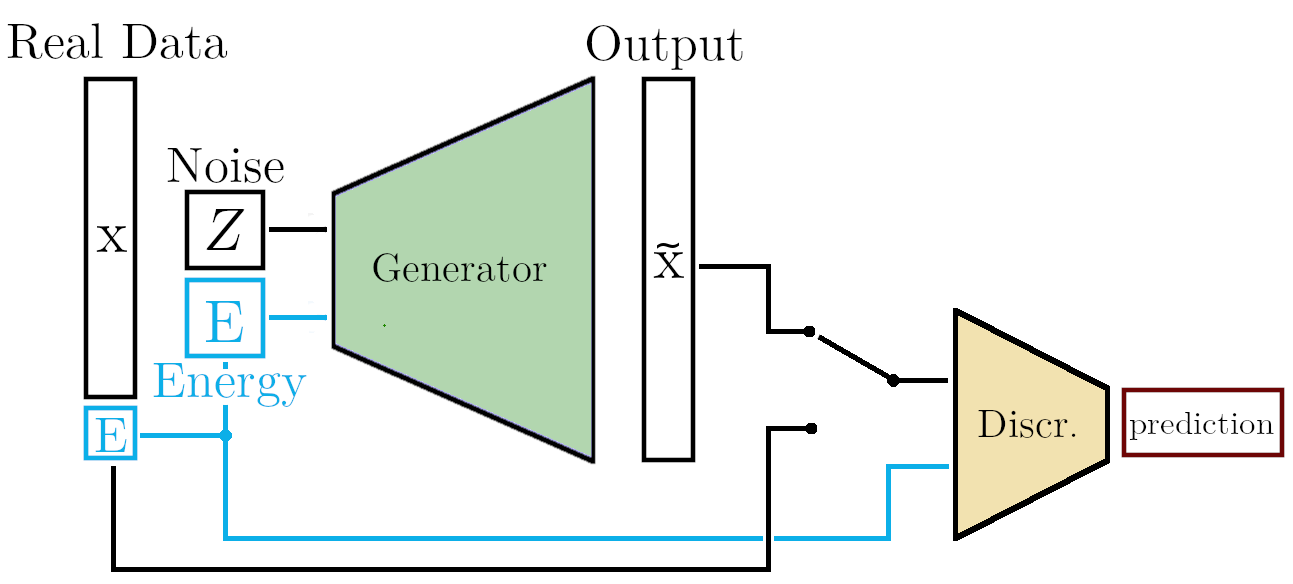}
    
    \vspace{0.3cm}
    
    \includegraphics[width=0.48\textwidth]{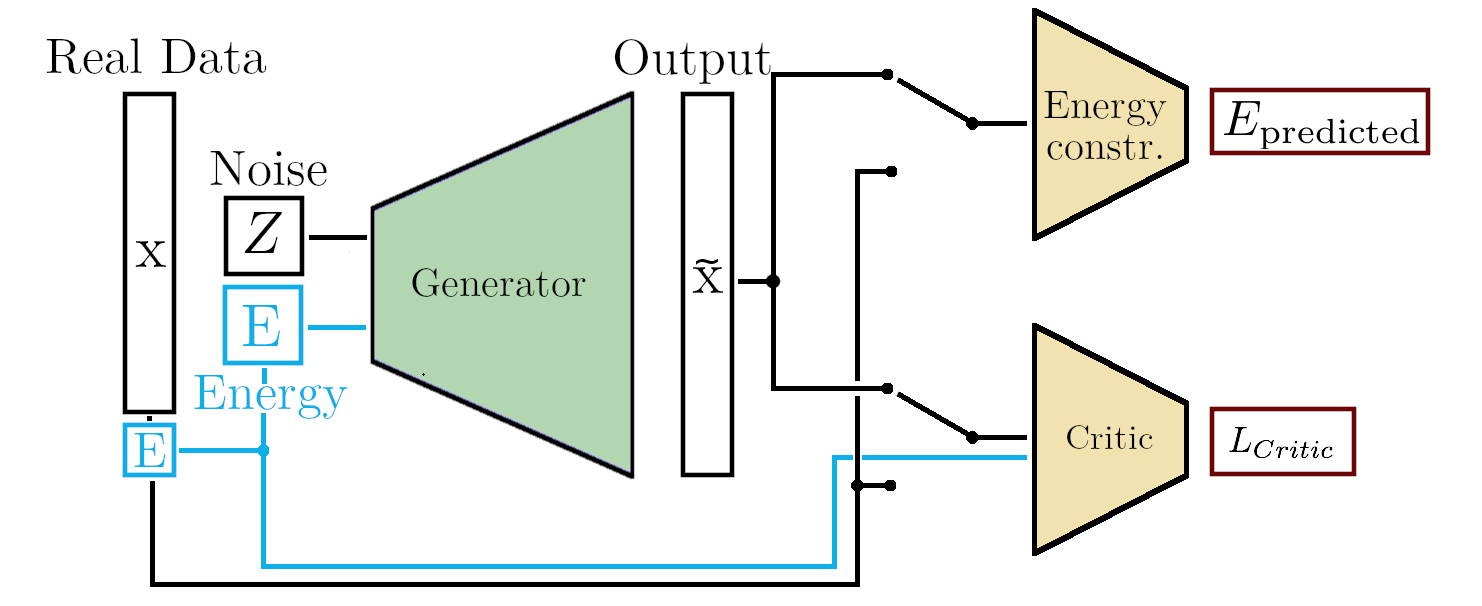}
    \caption{Overview of the GAN (top) and WGAN (bottom) architectures.
    The blue line shows where the true energy is used as an input. 
    The loss functions and feedback loops are explained in the text. 
    }
    \label{fig:VGAN}
\end{figure}

The GAN architecture was proposed in 2014~\cite{GAN} and had remarkable success
in a number of generative tasks. It introduces generative models by an adversarial process, in which a generator $G$ competes against an adversary (or discriminator) $D$. The goal of this framework is to train $G$ in order to generate samples $\widetilde{x}=G(z)$ out of noise $z$, which are indistinguishable from real samples $x$. The adversary network $D$ is trained to maximize the probability of correctly classifying whether or not a sample came from real data
using the binary cross-entropy.
The generator, on the other hand, is trained to fool the adversary $D$.
This is represented by the loss function as 
\begin{equation}
\label{eqn:JS}
\begin{split}
    L = \min_G \max_D \mathbb{E}[ & \log D(x)] + \mathbb{E}[\log (1-D(G(z)))],
\end{split}
\end{equation}
and a schematic of the GAN training is provided in Fig.~\ref{fig:VGAN}~(top). 

%

For practical applications, the GAN needs to simulate showers of a specific energy. To this end, 
we parameterise generator and discriminator as functions of the photon energy $E$~\cite{Parametrised}. In general, we attempted to minimally modify
the CaloGAN formulation~\cite{CaloGAN3} to work with the present dataset.

The original formulation of a GAN produces a generator that
minimizes the Jensen-Shannon divergence between true and generated data.
In general, the training of GANs is known to be technically challenging and subject to
instabilities~\cite{impGAN}. Recent progress on generative models 
improves upon this by modifying the learning objective.



\subsection{Wasserstein-GAN}

\label{sec:WGAN}

One alternative to classical GAN training is to use the Wasserstein-1 distance, also known as earth mover's distance, as a loss function. This distance evaluates dissimilarity between two multi-dimensional distributions and informally gives the cost expectation for moving a mass of probability along optimal transportation paths~\cite{optimal_transport}.  Using the Kantorovich-Rubinstein duality, the Wasserstein loss can be calculated as
\begin{equation}
    \label{eqn:wloss}
    L = \textrm{sup}_{f\in \textrm{Lip}_1}\{\mathbb{E}[f(x)] - \mathbb{E}[f(\tilde{x})]\}.
\end{equation}
The supremum is over all 1-Lipschitz functions $f$, which is approximated by a discriminator network $D$ during the adversarial training. This discriminator is called \textit{critic} since it is trained to estimate the Wasserstein distance between real and generated images.


In order to enforce the 1-Lipschitz constraint on the critic~\cite{WGAN2}, a gradient penalty term
should be added to (\ref{eqn:wloss}), yielding the critic loss function:
\begin{equation}
    \label{eqn:wgan_gp}
    \begin{split}
    L_{\textrm{Critic}} = \mathbb{E}[D(G(z))] & - \mathbb{E}[D(x)]  \\
    & + \lambda\,\, \mathbb{E}[(\parallel\nabla_{\hat{x}}D(\hat{x})\parallel_2 - 1)^2 ],
    \end{split}
\end{equation}
where $\lambda$ is a hyper parameter for scaling the gradient penalty. The term $\hat{x}$ is a mixture of real data $x$ and generated $G(z)$ showers. Following~\cite{WGAN2}, it is sampled uniformly along linear interpolations between $x$ and $G(z)$. 

Finally, we again need to ensure that generated showers accurately resemble photons of the 
requested energy. We achieve this by parametrising the generator and critic networks in $E$ and
by adding a constrainer~\cite{ErdmannWGAN2} network $a$.
The loss function for the generator then reads:
\begin{equation}
    \label{eqn:constrE}
    \begin{split}
    L_{\textrm{Generator}} = & -\mathbb{E}[D(\tilde{x},E)] \\
    & + \kappa\cdot \mathbb{E}[\left| (a(\tilde{x}) - E)^2 - (a(x) - E)^2 \right| ],
    \end{split}
\end{equation}
where $\tilde{x}$ are generated showers and $\kappa$ is the relative strength of the conditioning term. 
This combined network is illustrated in Fig.~\ref{fig:VGAN}.
The constrainer network is trained solely on the Geant4 showers; its weights are fixed during the generator training.  We use the mean absolute error (L1) as loss~\footnote{Using L1 loss here gives better performance than L2, as L2 seems to introduce too large a penalisation for the occasionally expected outliers in the total energy sum due to the finite calorimeter resolution.}: 
\begin{equation}
    \label{eqn:constrE_sup}
    L_\textrm{Constrainer} = \left|E - a(x)\right|.
\end{equation}


\subsection{Bounded Information Bottleneck-Autoencoder}

\label{sec:BIBAE}

\begin{figure*}[h]
    \centering
    \includegraphics[width=0.8\textwidth]{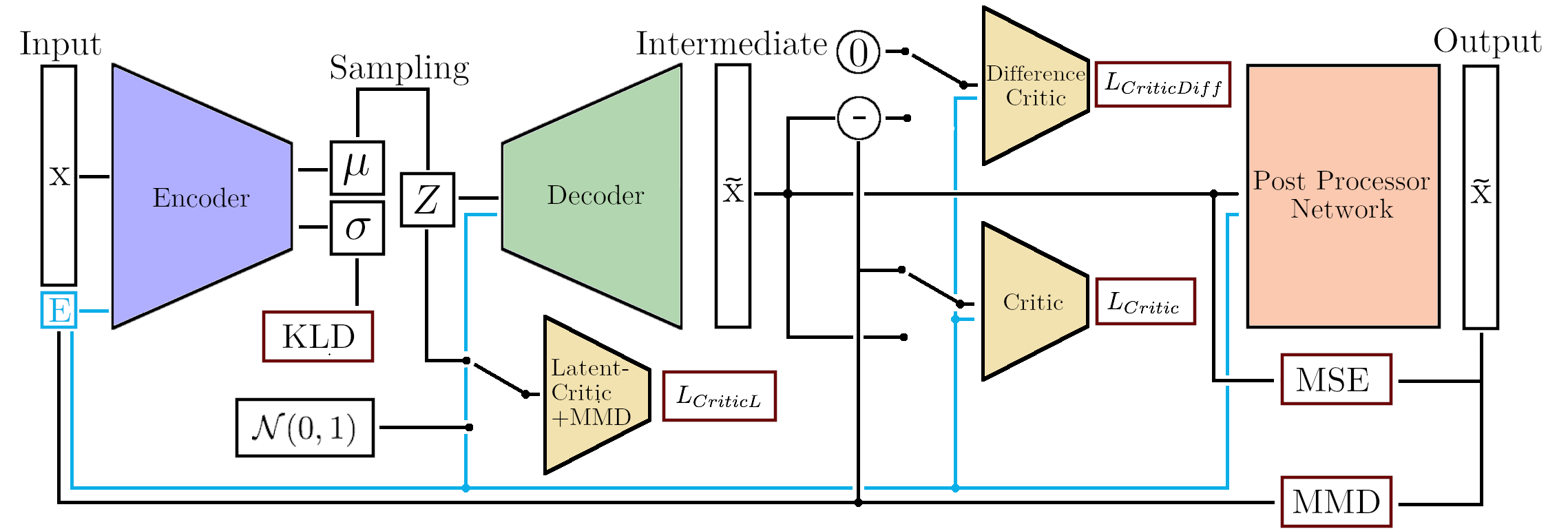}
    \caption{Diagram of the BIB-AE architecture, including the additional MMD term defined in Sec.~\ref{sec:MMD}
    and the Post Processor Network defined in Sec.~\ref{sec:PP}. The blue line shows where the true energy is used as an input. 
    The loss functions and feedback loops are explained in the text. }
    \label{fig:BAE_PP}
\end{figure*}

Autoencoder architectures map input to output data via a latent space. 
Using a structured latent space allows for later sampling and thereby generation of new data.
The BIB-AE~\cite{BIB-AE} architecture was introduced as a theoretical overarching generative model.
Most commonly employed generative models --- e.g. GAN \cite{GAN}, VAE \cite{VAE}, and adversarial autoencoder (AAE)~\cite{AAE} --- can be seen as different subsets of the BIB-AE.
This leads to better control over the latent space distributions and promises better generative performance and interpretability. 
In the following, we focus on the practical advantage gained from utilizing the individual BIB-AE components and refer to the original publication~\cite{BIB-AE} for an
information-theoretical discussion.

As it is an overarching model, an instructive way for describing the base BIB-AE framework is by taking a VAE and expanding upon it. A default VAE consist of four general components: an encoder, a decoder, a latent-space regularized by the Kullback--Leibler divergence (KLD), and an $L_N$-norm to determine the difference between the original and the reconstructed data. These components
are all present as well in the BIB-AE setup.
Additionally, one introduces a GAN-like adversarial network, trained to distinguish between real and reconstructed data, as well as a sampling based method of regularizing the latent space, such as another adversarial network or a maximum mean discrepancy (MMD, as described in the next section) term. In total this adds up to four loss terms: The KLD on the latent space, the sampling regularization on the latent space, the $L_N$-norm on the reconstructed samples and the adversary on the reconstructed samples. The guiding principle behind this is that the two latent space and the two reconstruction losses complement each other and, in combination, allow the network to learn a more detailed description of the data. Specifically looking at the two reconstruction terms we have, on the one hand, the adversarial network: from tests on utilizing GANs for shower generation we know that such adversarial networks are uniquely qualified to teach a generator to reproduce realistic looking individual showers. On the other hand, we have the $L_N$-norm: while our trials with pure VAE setups have shown that $L_N$-norms have great difficulty capturing the finer structures of the electromagnetic showers, an $L_N$-norm also forces the encoder-decoder structure to have an expressive latent space, as the original images could not be reconstructed without any latent space information. Therefore, the adversarial network forces the individual images to look realistic, while the $L_N$-norm forces latent space utilization, thereby improving how well the overall properties of the data set are reproduced. The latent space loss terms have a similar interaction. Here the KLD term regularizes our complete latent space by reducing the difference between the average latent space distribution and a normal Gaussian. The KLD is, however, largely blind to the shape of the individual latent space dimensions, as it only cares about the average. The sampling based latent space regularization term fills this niche by looking at every latent space dimension individually.

Our specific implementation of the BIB-AE framework is shown in Fig.~\ref{fig:BAE_PP}. For our sampling based latent regularization we use both an adversary and an MMD term. The adversaries are implemented as critics trained with gradient penalty, similar to the WGAN approach. The main difference in our setup compared to the one described in~\cite{BIB-AE} is that we replaced the $L_N$-norm with a third critic, trained to minimize the difference between input and reconstruction. We chose this because we found that using the $L_N$-norm to compare the input and the reconstructed output resulted in smeared out images.

For the precise implementation of the loss functions we define the encoder network $N$, the decoder network $D$, the latent critic $C_L$, the critic network $C$, 
and the difference critic $C_D$. 
The loss function for the latent critic $C_L$ is given by
\begin{equation}
    \label{eqn:bibae_crit_L}
    \begin{split}
    L_{C_{L}} = & \mathbb{E}[C_L(N_{E}(x))] - \mathbb{E}[C_L(\mathcal{N}(0,1))]  \\
    & + \lambda\,\, \mathbb{E}[(\parallel\nabla_{\hat{x}}C_L(\hat{x})\parallel_2 - 1)^2 ].
    \end{split}
\end{equation}
Here $\hat{x}$ is a mixture of the encoded input image $N(x)$ and samples from a normal distribution $\mathcal{N}(0,1))$ and the $E$ subscript indicates that the network receives the photon energy label as an input.
The loss function for the main critic $C$ is given by
\begin{equation}
    \label{eqn:bibae_crit}
    \begin{split}
    L_{C} = & \mathbb{E}[C_{E}(D_{E}(N_{E}(x)))] - \mathbb{E}[C_{E}(x)]  \\
    & + \lambda\,\, \mathbb{E}[(\parallel\nabla_{\hat{x}}C_{E}(\hat{x})\parallel_2 - 1)^2 ].
    \end{split}
\end{equation}
Where $\hat{x}$ is a mixture of the reconstructed image $D(N(x))$ and the original images $x$.
Finally, the loss function for the difference critic $C_D$ is given by
\begin{equation}
    \label{eqn:bibae_crit_D}
    \begin{split}
    L_{C_D} = \mathbb{E} & [C_{D,E}(D_{E}(N_{E}(x)) - x)]  - \mathbb{E}[C_{D,E}(x - x=0)]  \\
    & + \lambda\,\, \mathbb{E}[(\parallel\nabla_{\hat{x}}C_{D,E}(\hat{x})\parallel_2 - 1)^2 ].
    \end{split}
\end{equation}
Where $\hat{x}$ is a mixture of the difference $D(N(x)) - x$ and the difference $x - x=0$.
With different $\beta$ factors 
giving the relative weights for the individual loss terms,
the combined loss for the encoder and decoder parts of the BIB-AE can be expressed as:
\begin{equation}
    \label{eqn:BIBAE}
    \begin{split}
    L_{\textrm{BIB-AE}} = & - \beta_{C_L} \cdot \mathbb{E}[C_{L}(N_{E}(x))] \\
    & - \beta_{C} \cdot \mathbb{E}[C_{E}(D_{E}(N_{E}(x)))] \\
    & - \beta_{C_D} \cdot \mathbb{E}[C_{D,E}(D_{E}(N_{E}(x)) - x)] \\
    & + \beta_{\textrm{KLD}} \cdot \textrm{KLD}(N_{E}(x)) \\
    & + \beta_{\textrm{MMD}} \cdot \textrm{MMD}(N_{E}(x),\mathcal{N}(0,1))).
    \end{split}
\end{equation}

\subsection{Maximum Mean Discrepancy}
\begin{figure*}[h!]
    \centering
    \includegraphics[width=0.286\textwidth]{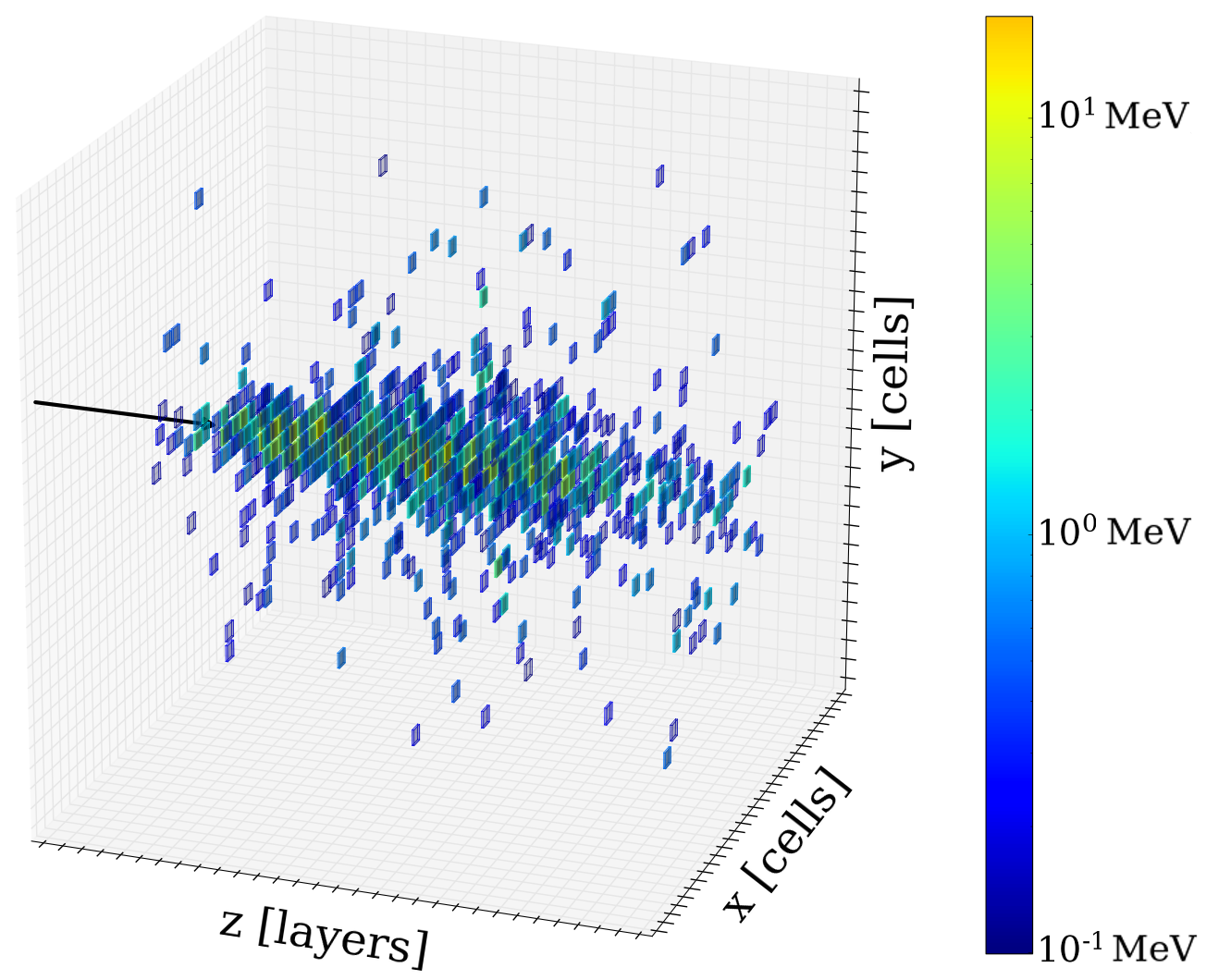}
    \includegraphics[width=0.22\textwidth]{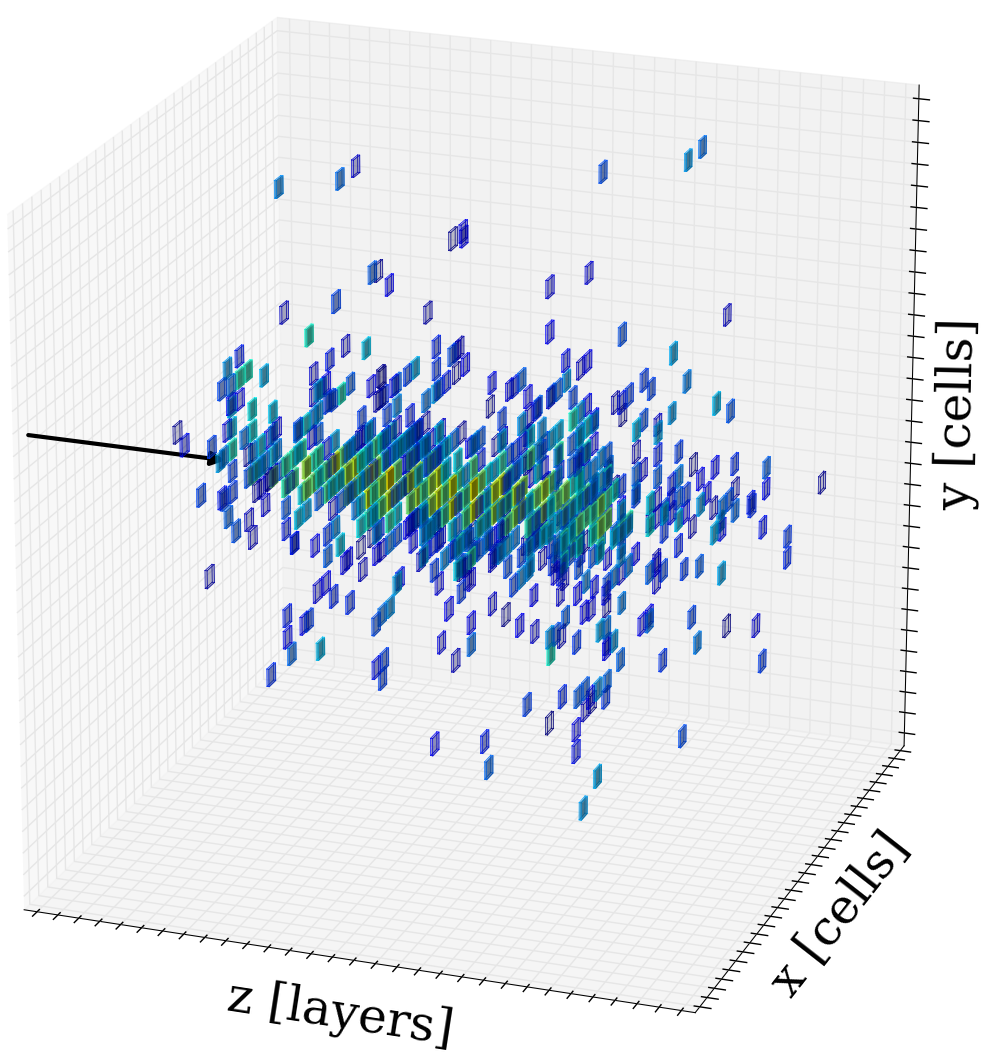}
    \includegraphics[width=0.22\textwidth]{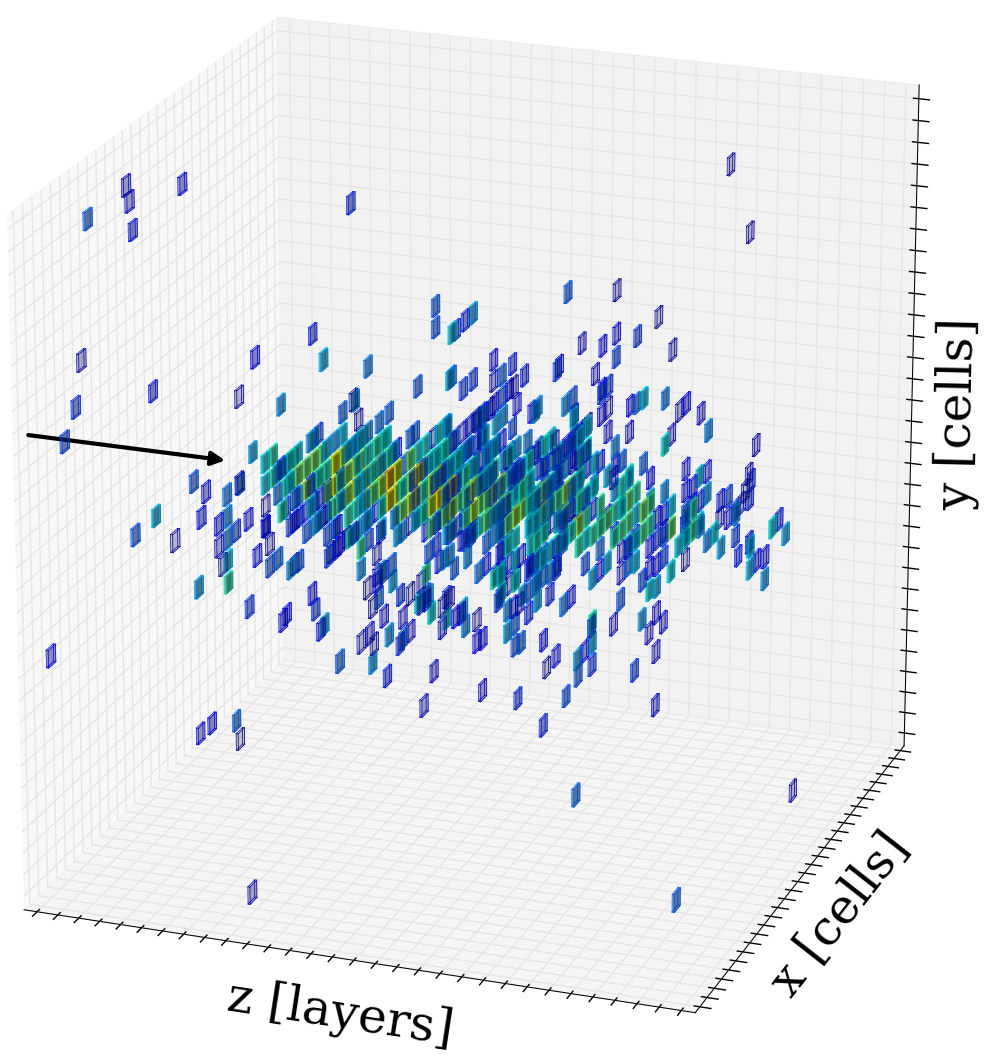}
    \includegraphics[width=0.22\textwidth]{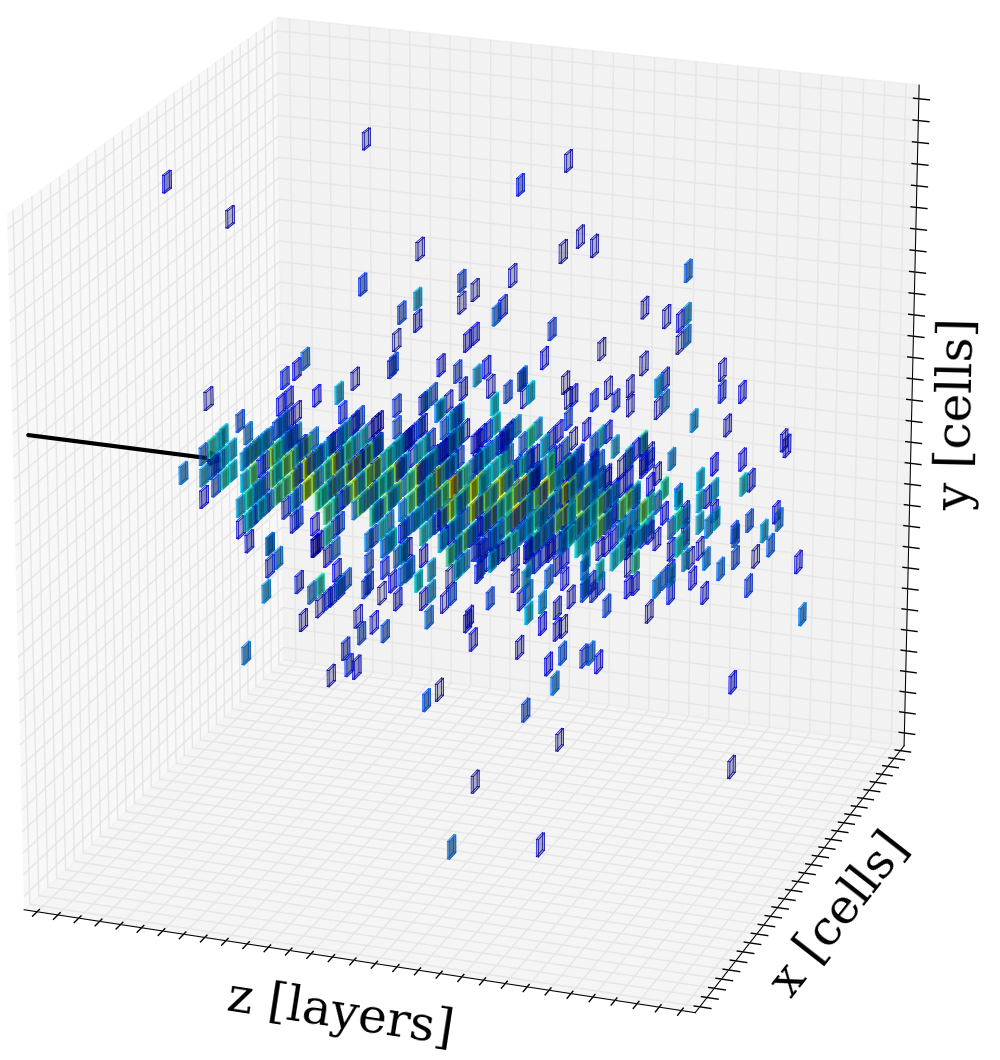}
    \caption{
    Examples of individual 50~GeV photon showers generated by Geant4 (left), the GAN (center left), WGAN (center right), and BIB-AE (right)
    architectures. Colors encode the deposited energy per cell.}
    \label{fig:results_3dplots}
\end{figure*}

\label{sec:MMD}

One major challenge in generating realistic photon showers is the spectrum of the individual cell energies, which is shown in  Fig.~\ref{fig:results_diffdist1}~(left)
in Section~\ref{results}.
The real spectrum shows an edge around the energy that a minimal ionizing particle (MIP) would deposit. Since the well-defined energy deposition of a MIP is often used to calibrate a calorimeter, we cannot simply ignore it. However, we found that purely adversarial based methods tend to smooth out this and other similar low energy features, an observation in line with other efforts to use generative networks for shower simulation~\cite{ErdmannWGAN2}. A way of dealing with this is using MMD~\cite{MMD_base} to compare and minimize the distance between the real $(D_{R})$ and fake $(D_{F})$ hit-energy distributions:
\begin{align}
\label{eq:mmd}
\begin{split}
    \textrm{MMD}(D_{R}, D_{F}) = & \langle k(x,x') \rangle + \langle k(y,y') \rangle  \\
    & - 2\langle k(x,y) \rangle ,
\end{split}
\end{align}
where $x$ and $y$ are samples drawn from $D_{R}$ and $D_{F}$ respectively and $k$ is any positive definite kernel function. MMD based losses have previously been used in the generation of LHC events~\cite{LHCGAN}. 

A naive implementation of the MMD would be to compare every pixel value from a real shower with every value from a generated shower. This approach is however not feasible since it would involve computing Equation (\ref{eq:mmd}) approximately $(30^{3})^{2}$ times for each shower.
To make the MMD calculation tractable, we introduce a novel version of the MMD, termed Sorted-Kernel-MMD. We first sort both, real and generated, hit-energies in descending order, and then  take the $n$ highest fake energies and compare them to the $n$ highest real energies. Following this we move the $n$-sized comparison window by $m$ and recompute the MMD. This process is repeated $\frac{N}{m}$-times, where $N$ is the total number of pixels one wants to compare. The advantage of this approach is two-fold, for one the number of computations is linear in $N$, 
as opposed to the naive implementation
which shows quadratic behavior. 
The second advantage is that energies will only be compared to similar values, thereby incentivising the model to fine-tune 
the energy. 
Specifically, the values m=25, and n=100 are used and we chose N=2000, as this is approximately the maximum occupancy observed in our training data before any low energy cutoffs.
In our experiments, adding this MMD term with the kernel function 
\begin{equation}
    k(x,x^{\prime}) = e^{- \alpha (x^{2} + x^{\prime 2} - 2 x x^{\prime})} 
\end{equation} 
with $\alpha = 200$ to the loss term of either a GAN or a BIB-AE fixes the per-cell hit energy spectrum to be near identical to the training data. This however comes at a price, as the additional pixels with the energies used to fix the spectrum are often placed in unphysical locations, specifically at the edges of the $30 \times 30 \times 30$ cube. 

\subsection{Post Processing}

\label{sec:PP}

In the previous section we found that using an MMD term in the loss function represents a trade off between correctly reproducing either the hit energy spectrum or the shower shape. 
To solve this, we split the problem into two networks that are applied consecutively but trained with different loss functions. The first network is a GAN or BIB-AE trained without the MMD term. This produces showers with correct shapes, but an incorrect hit-energy spectrum. The second network then takes these showers as its input and applies a series of convolutions with kernel size one. Therefore this second network can only modify the values of existing pixels, but not easily add or remove pixels. This second network, here called Post Processor Network, is trained using only the MMD term to fix the hit energy spectrum, and the mean squared error (MSE) between the input and output images, ensuring the change from the Post Processor Network is as minimal as possible. 
\done{Need to address author's comment on equivalence to simple f(E) vs. convolutions -> appendix}



\section{Results}

\label{results}

In the following we present the ability of our generative models to accurately predict a number
of per-shower variables as well as global observables and analyse the achievable gain in computing performance. 
We include our implementation of a simple GAN~(Sec.~\ref{sec:GAN}), 
a WGAN with additional energy constrainer (Sec.~\ref{sec:WGAN}),
and a BIB-AE with energy-MMD and post processing (Secs.~\ref{sec:BIBAE},~\ref{sec:MMD} and ~\ref{sec:PP}).
A detailed discussion of the architectures and training hyper parameters can be found in Appendix~\ref{hyper}.
All architectures are trained on the same sample of 950k Geant4 showers.
Tests are either shown
for the full momentum range (labeled \textit{full spectrum}) or for specific shower energies (labeled 
with the incident photon energy in GeV).

\subsection{Physics Performance}

\begin{figure*}[t]
    \centering
    \includegraphics[width=0.42\textwidth]{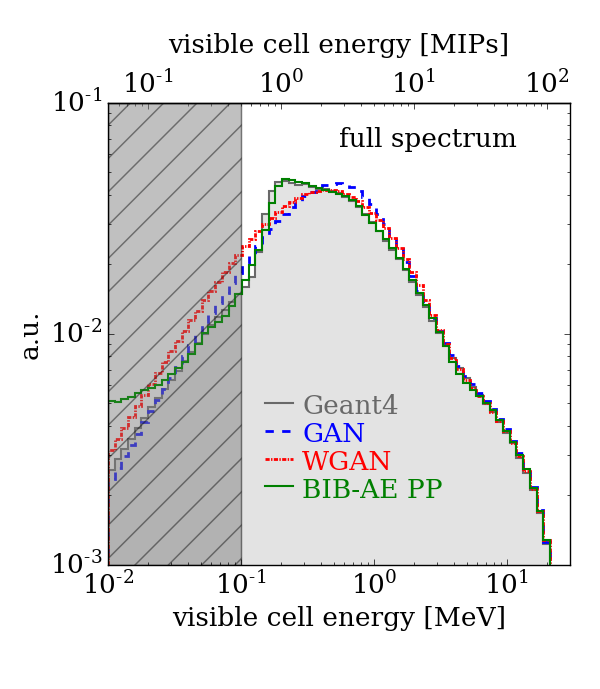}
    \includegraphics[width=0.42\textwidth]{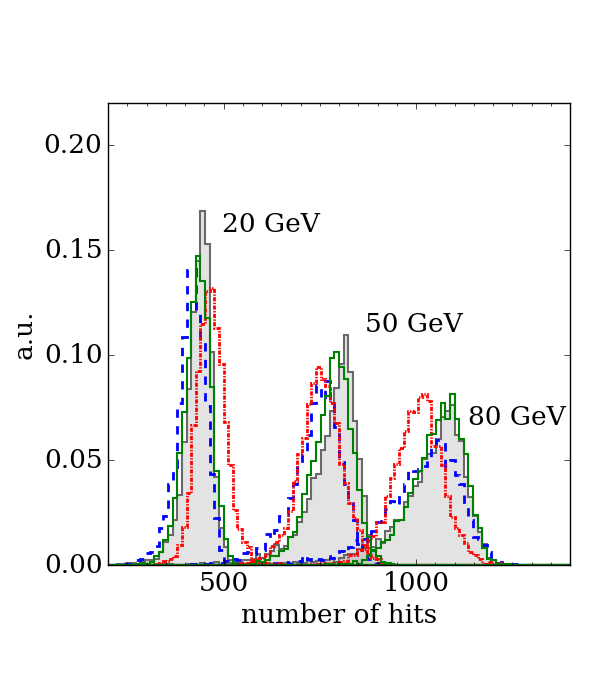}
    \caption{Differential distributions comparing the per-cell energy (left) and the number of hits above 0.1 MeV (right) between Geant4 and the different generative models.
    Shown are Geant4 (grey, filled), our GAN setup (blue, dashed), our WGAN (red, dotted) and the BIB-AE (green, solid).
    The energy per-cell is measured in MeV for the bottom axis and in multiples of the expected energy deposit of a minimum ionizing particle (MIP) for
    the top axis.
    }
    \label{fig:results_diffdist1}
\end{figure*}

We first verify in Fig.~\ref{fig:results_3dplots} that the showers generated by
all network architectures visually appear to be acceptable compared to Geant4.
Were we attempting to generate \textit{cute cat pictures}, our work would be done already at this point. 
Alas, these shower images are eventually to be used as realistic substitutes in physics analyses so we 
need to pay careful attention to relevant differential distributions and correlations.

\begin{figure*}[h]
    \centering
    \includegraphics[width=0.3\textwidth]{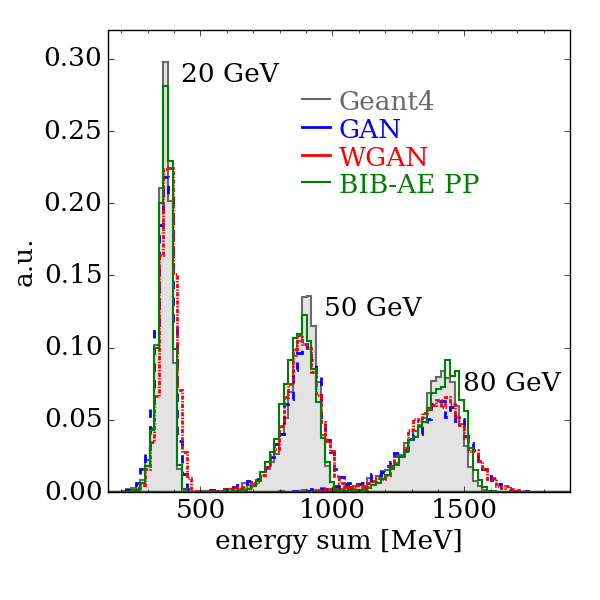}
    \includegraphics[width=0.3\textwidth]{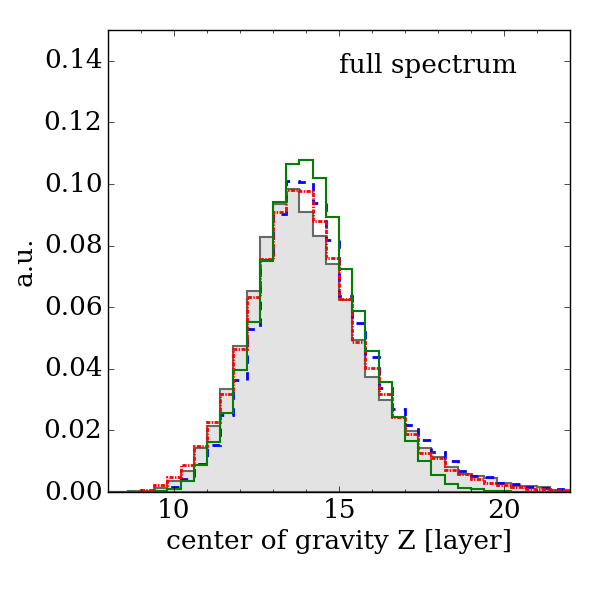}\\
    \includegraphics[width=0.3\textwidth]{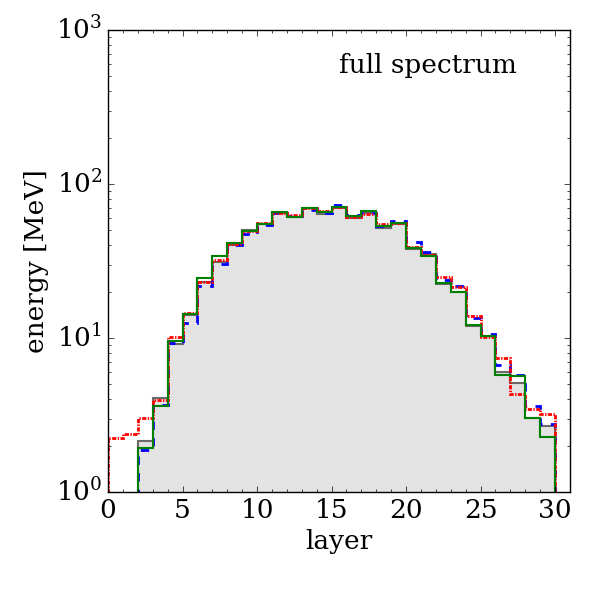}
    \includegraphics[width=0.3\textwidth]{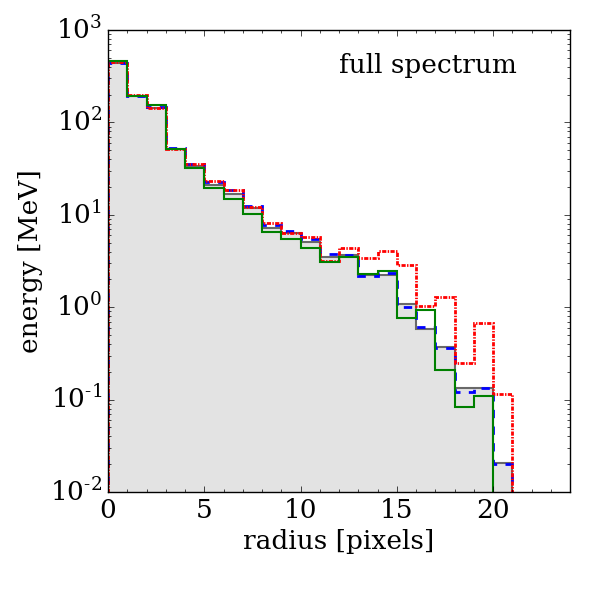}
    \caption{Additional differential distributions comparing physical observables between Geant4 and the different generative models.
    Shown are Geant4 (grey, filled), our GAN setup (blue, dashed), our WGAN (red, dotted) and the BIB-AE with Post Processing (green, solid).
    }
    \label{fig:results_diffdist2}
\end{figure*}

\begin{figure*}[t]
    \centering
    \includegraphics[width=0.34\textwidth]{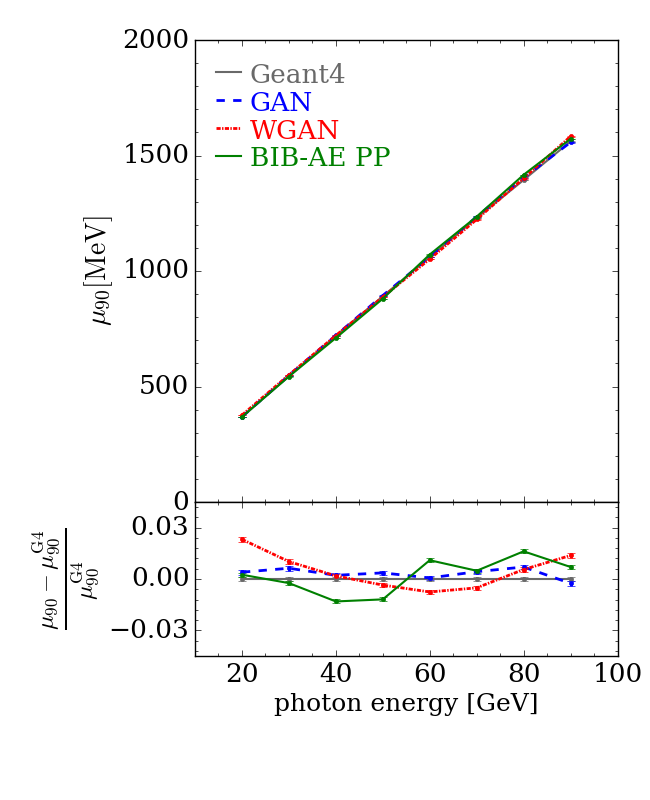}
    \includegraphics[width=0.34\textwidth]{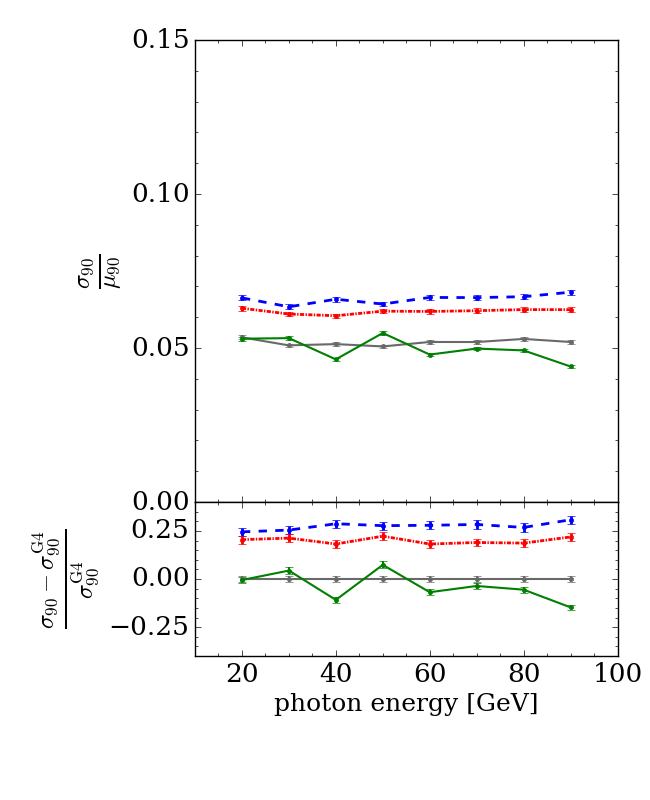}
    \caption{Plot of mean ($\mu_{90}$, left) and relative width ($\sigma_{90}/\mu_{90}$, right)  of the energy deposited in the calorimeter for various incident particle energies. In order to avoid edge effects, the phase space boundary regions of 10 and 100 GeV are removed for the response and resolution studies.
    In the bottom panels, the relative offset of these quantities with respect to the Geant4 simulation is shown.}
    \label{fig:results_mean}
\end{figure*}

In Figure~\ref{fig:results_diffdist1} a comparison between two differential distributions for 
all studied architectures and Geant4 is shown. The left plot compares the per-cell hit-energy spectrum
averaged over showers for the full spectrum of photon energies. 
We observe that while the high-energy hits are well described by all generative models, both GAN and WGAN fail to capture the bump around 
$0.2$~MeV. The BIB-AE is able to replicate this feature thanks to the Post Processor Network.\footnote{We studied applying post processing to the WGAN architecture as well. This is discussed in Section~\ref{subsec:WGAN_PP}.} This energy corresponds to the most probable energy loss of a MIP passing a silicon sensor of the ILD Si-W ECal at perpendicular incident angle. Since this is a well-defined energy, it can be used in highly granular calorimeters for the equalisation of the cell response as well as for setting an absolute energy scale. 
It also leads to a sharp rise in the spectrum, as lower energies can only be deposited by ionizing particles that pass only a fraction of the thickness at the edges of sensitive cells or that are stopped.
The region below half a MIP, corresponding to around 0.1~MeV,  is shaded in dark grey. These cell energies are very small and therefore 
will be discarded in a realistic calorimeter, as their signal to noise ratio is too low. 
For the following discussion cell energies below 0.1~MeV will therefore not be considered and only cells above this cut-off are included in all other performance plots and distributions.

Next, the plot on the right shows the number of hits for three discrete photon energies (20 GeV, 50 GeV, and 80 GeV). 
Here, the GAN and WGAN setups slightly underestimate the total number of hits, while the BIB-AE accurately models the mean and width 
of the distribution. This behavior can be traced back to the left plot. Since we apply a cutoff removing hits below $0.1\ \textrm{MeV}$, a model that does not correctly reproduce the hit-energy spectrum around the cut-off will have difficulties correctly describing the number of hits. 

Additional distributions are shown in Fig.~\ref{fig:results_diffdist2}. The top left depicts the visible energy distribution for the same three discrete photon energies. 
Both, the shape, center and width of the peak are well reproduced for all models. 
Due to the sampling nature of the calorimeter under study, the visible energy is of course much lower
than the incoming photons' energy.
    
In the top right and bottom two plots we compare the spatial properties of the generated showers. 
First, on the top right, the position of the center of gravity along the z axis
is shown. The Geant4 distribution is well modelled by the GANs, however there are slight deviations for the BIB-AE.
A detailed investigation of this discrepancy showed that the z axis center of gravity is largely encoded in a single
latent space variable. A mismatch between the observed latent distribution for real samples and the normal distribution 
drawn from when generating new samples directly translates into the observed difference. Sampling from a modified
distribution would remove the problem. 

Finally, the two plots on the bottom show the longitudinal and radial energy distributions.
We see that while all models are able the reproduce the bulk of the distributions very well, deviations for the WGAN appear around the edges. 


We next test how well the relation of visible energy to the incident photon energy is reproduced. To this end we use a Geant4 sample where we simulated photons at discrete energies ranging from 20 to 90 GeV in 10 GeV steps. We then use our models to generate showers for these energies and calculate the 
mean and root-mean-square of the $90\%$~core of the distribution, labeled $\mu_{90}$ and $\sigma_{90}$ respectively, 
for all sets of showers. The results are shown in Fig.~\ref{fig:results_mean}. Overall the mean (left) is correctly modelled, showing only deviations in the order of one to two percent. 
The relative width, $\sigma_{90}/\mu_{90}$ (right) looks worse: GAN and WGAN overestimate the Geant4 value at all energies.
While the BIB-AE on average correctly models the width, it still shows deviations of up to ten percent
at high energies. Note that the width cannot be interpreted as energy resolution of the calorimeter due to the two different absorber thicknesses used in the ECal, requiring different calibrations.

\begin{figure*}[h]
    \centering
    \includegraphics[width=0.45\textwidth]{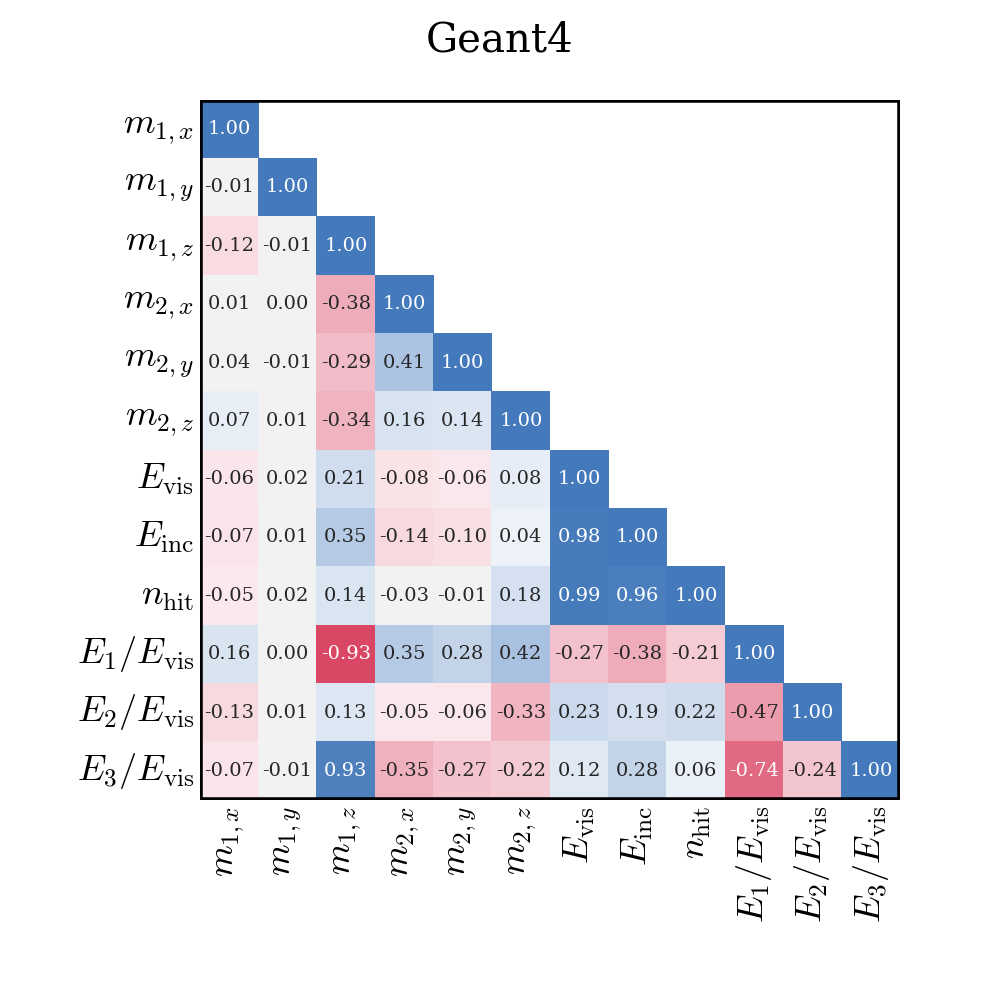}
    \includegraphics[width=0.45\textwidth]{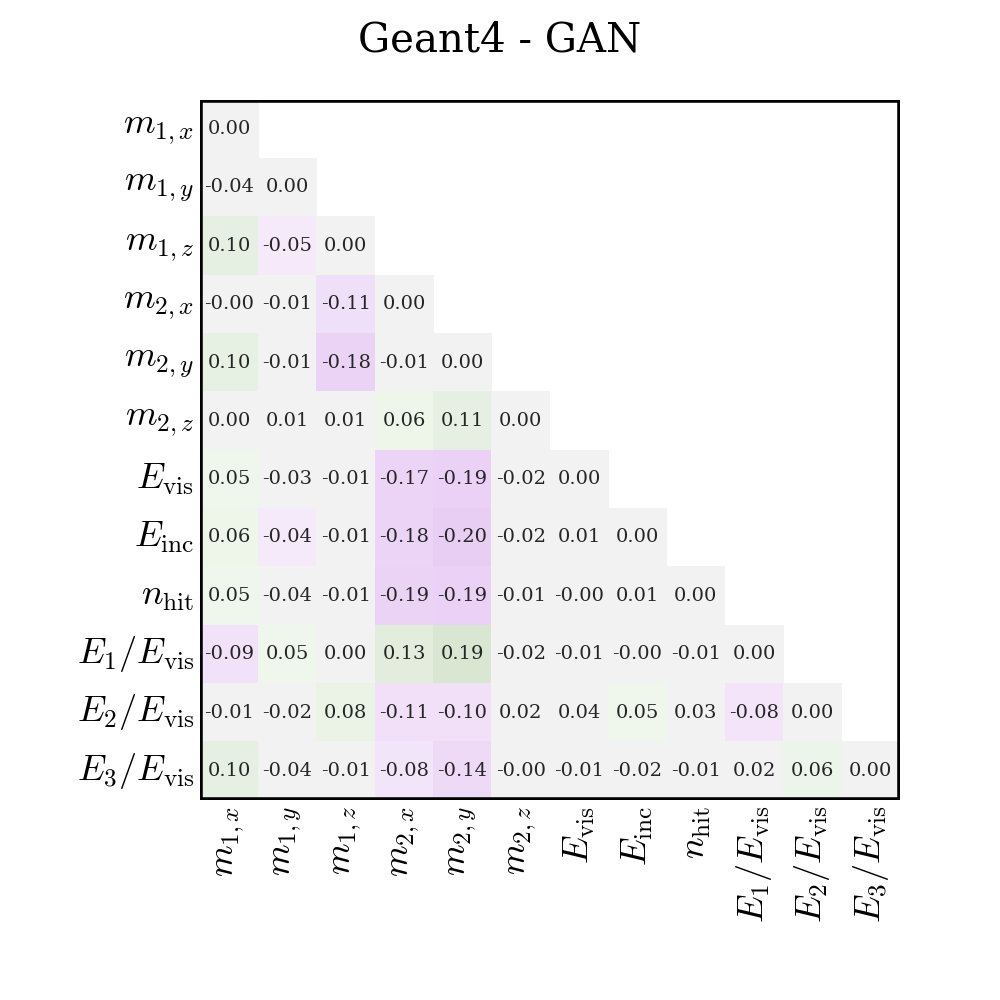}    
    \includegraphics[width=0.45\textwidth]{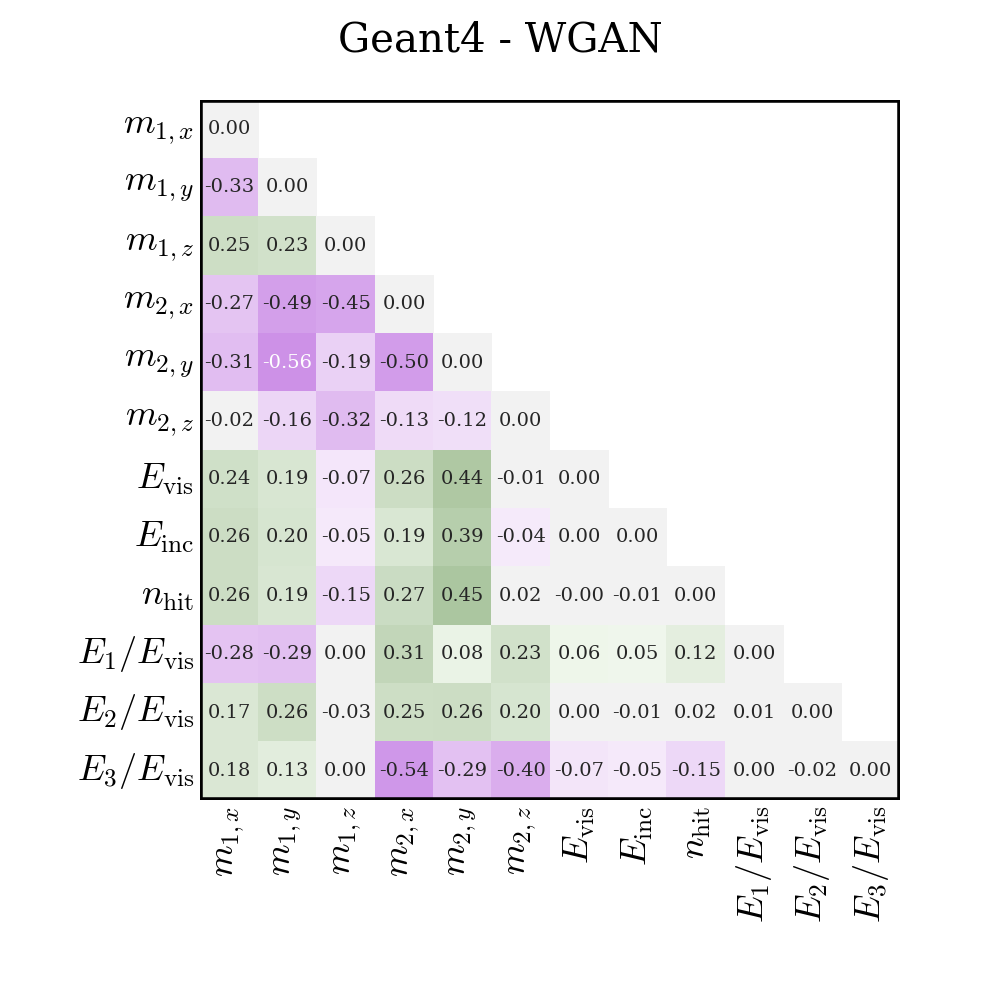}
    \includegraphics[width=0.45\textwidth]{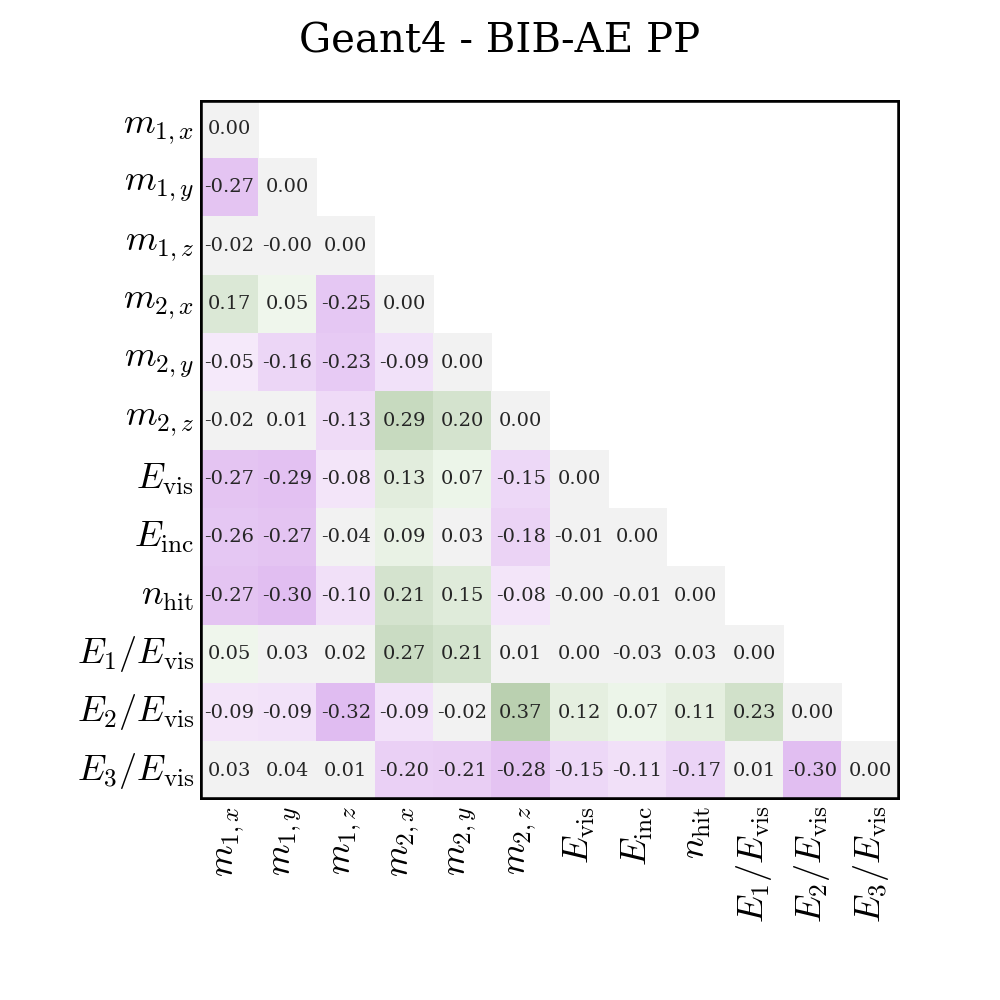}
    \caption{Linear correlation coefficients between various quantities described in the text in Geant4 (top left). 
    Difference between these correlations in Geant4 and GAN (top right),
    Geant4 and WGAN (bottom left), and Geant4 and BIB-AE with post processing (bottom right). The mean absolute differences compared to Geant4 are 0.058 for the GAN, 0.187 for the WGAN and 0.132 for the BIB-AE.
    }
    \label{fig:results_correlations}
\end{figure*}

\begin{figure*}[h]
    \centering
    \includegraphics[width=1.00\textwidth]{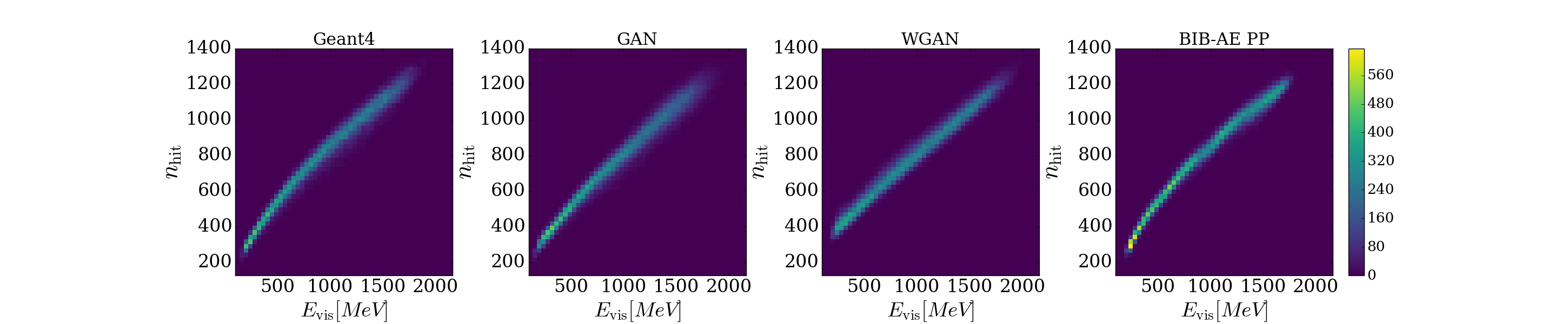}
    \includegraphics[width=1.00\textwidth]{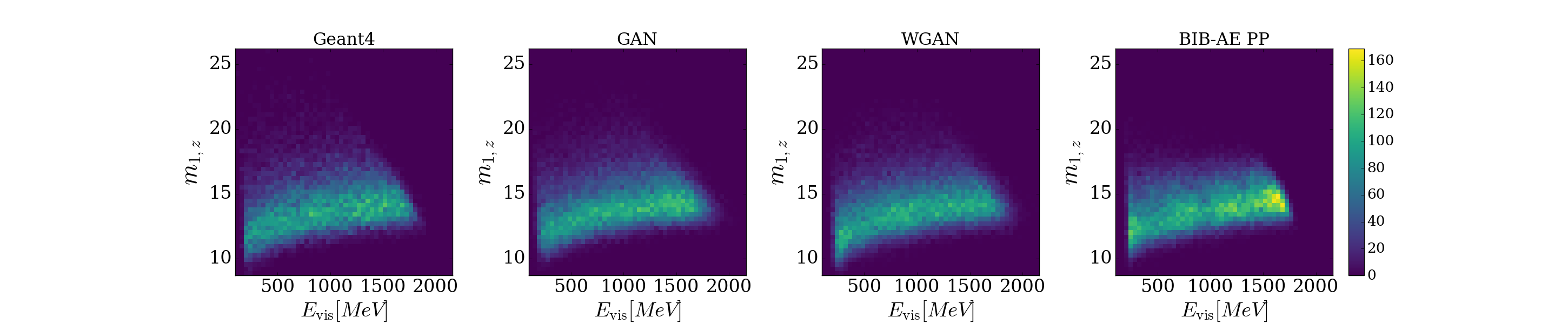}
    \caption{Scatter plot showing the correlations between visible energy and number of hits (top) and visible energy and center of gravity (bottom).
    }
    \label{fig:results_correlations_scatter}
\end{figure*}
Finally, we verify whether correlations between individual shower properties present in Geant4 are correctly reproduced by our generative setups. 
The properties chosen for this are: The first and second moments in x, y and z direction, labeled as $m_{1,x}$ through $m_{2,z}$, the visible energy deposited in the calorimeter $E_{\textrm{vis}}$, the energy of the simulated incident particle $E_{\textrm{inc}}$, the number of hits $n_{\textrm{hit}}$, and the ratio between the energy deposited in the 1st/2nd/3rd third of the calorimeter and the total visible energy, labeled $E_{1}/E_{\textrm{vis}}$ through $E_{3}/E_{\textrm{vis}}$. The results are shown in Fig.~\ref{fig:results_correlations}. 
The top left plot shows the correlations for Geant4 showers. 
We then present the difference to Geant4 for the GAN (top right), WGAN (bottom left),
and BIB-AE (bottom right). The smallest differences are observed for the GAN (absolute maximum difference of $0.2$), followed BIB-AE ($0.36$) and WGAN ($0.57$).

Fig.~\ref{fig:results_correlations_scatter} shows examples of 2D scatter plots: the number of hits and the visible energy (top row) as well as the center of gravity and the visible energy (bottom row). These allow us insight into the full correlations between these variables beyond the simple correlation coefficients. Similar to Fig.~\ref{fig:results_correlations} we see that the GAN matches the Geant4 correlations exceptionally well, while the WGAN and the BIB-AE display some slight correlation mis-matching. The discrepancy in the BIB-AE center of gravity and visible energy correlation can be traced back to the mismodelling of the center of gravity as seen in Fig.~\ref{fig:results_diffdist2}.

The distributions of physical observables shown above are expected to be the major factor for assessing the quality of a simulation tool.
While the correlations are also useful as they provide additional insight, our main focus when evaluating network performance are the phys-ics distributions.


\subsection{The importance of post processing}
\label{subsec:WGAN_PP}

\begin{figure*}[h]
    \centering
    \includegraphics[width=0.31\textwidth]{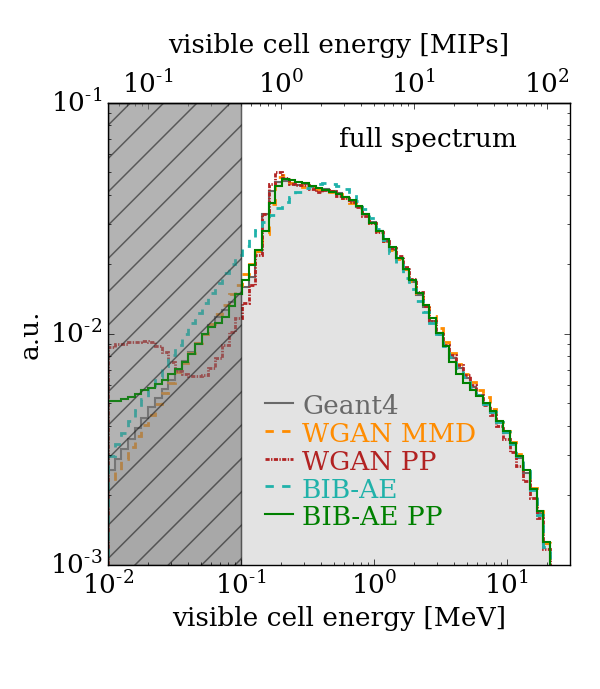}
    \includegraphics[width=0.31\textwidth]{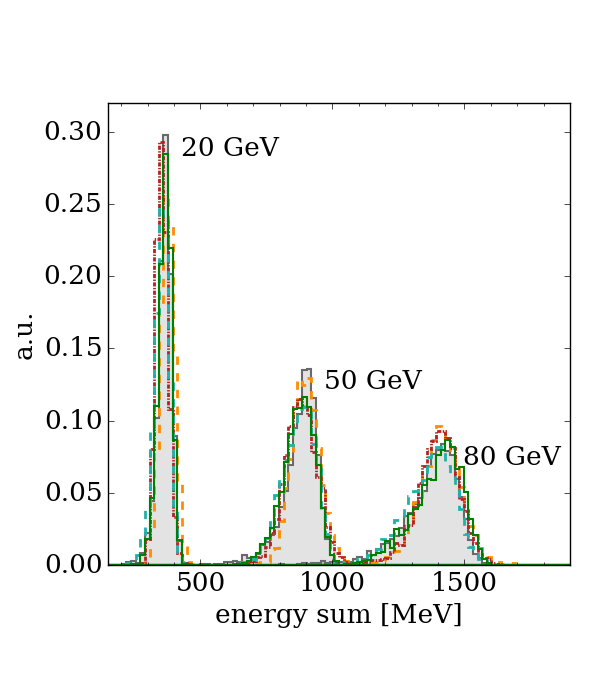}
    \includegraphics[width=0.31\textwidth]{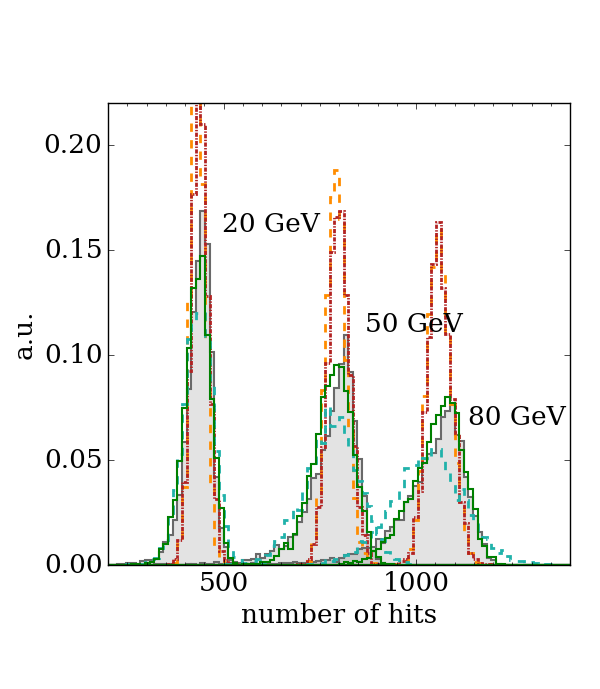}
    \includegraphics[width=0.31\textwidth]{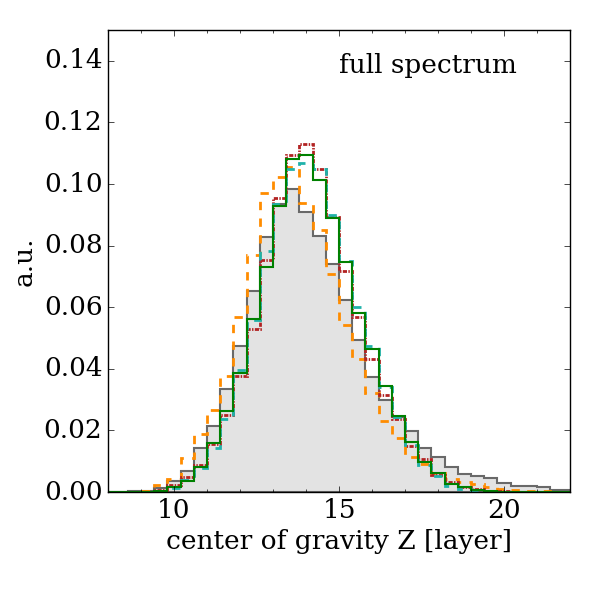}
    \includegraphics[width=0.31\textwidth]{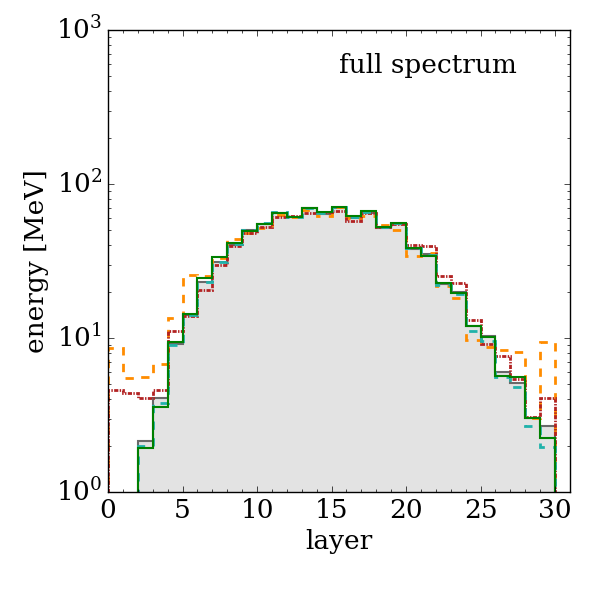}
    \includegraphics[width=0.31\textwidth]{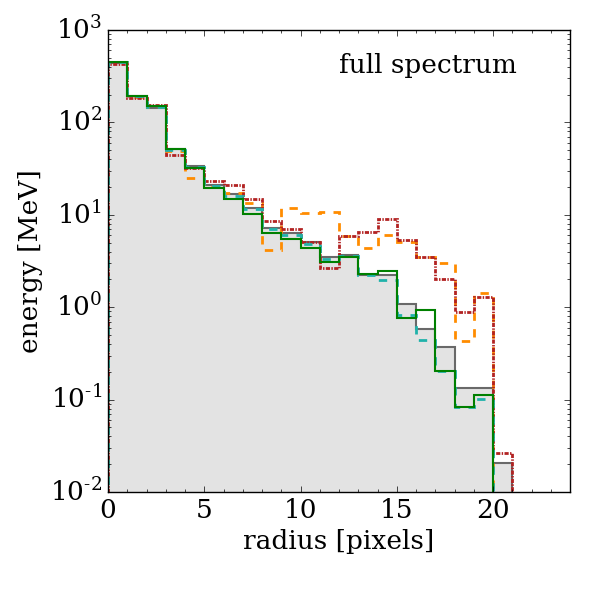}
    \caption{Differential distributions comparing physics quantities between Geant4 and the different generative models.
    The energy per-cell is measured in MeV for the bottom axis and in multiples of the expected energy deposit of a minimum ionizing particle (MIP) for
    the top axis.}
    \label{fig:appendix_results_diffdist}
\end{figure*}

In the previous section we demonstrated that our proposed architecture
--- the BIB-AE with a Post Processor Network --- 
achieved excellent performance in simulating important calorimeter observables.
In the following, we will dissect this improvement. To this end we compare a
WGAN trained with an additional simple MMD kernel (labelled WGAN MMD), 
a WGAN trained with the full post processing (labelled WGAN PP),
a BIB-AE without post processing (labelled BIB-AE) to
Geant4 and to the combined BIB-AE network including post processing (labelled BIB-AE PP) from the main text. We do not investigate a simple GAN with post processing as we expect it to exhibit largely the same behaviour as the WGAN.

In Fig.~\ref{fig:appendix_results_diffdist} we show the performance 
of these approaches. The top left panel of
Fig.~\ref{fig:appendix_results_diffdist}
demonstrates that removing post-processing 
from the BIB-AE leads to a smeared out MIP peak,
while adding
the simple MMD term or the more complex 
post processing to the WGAN result in good modelling of the 
per-cell hit energy spectrum. 
However, now this improvement comes at a price:
the distribution of the number of hits (top right) is too narrow
compared to Geant4 and 
the longitudinal (bottom center) and radial (bottom right)
energy profiles are described badly as additional 
energy is deposited at the edges of the shower.
Especially noticeable is the additional energy in the first and last layers. This would be problematic for standard reconstruction methods that rely on the precise position of the shower start and end. These energy deposits along the image edges are the main reason why the BIB-AE Post Processor is implemented as a separate network rather than integrated in the main decoder structure. The latter would require applying the MMD loss to the entire decoder, which in our test led to energy deposits similar to what can be seen in the WGAN MMD line.

While we were not able to improve the WGAN approach 
via post processing, we are not aware of fundamental reasons why a better 
performance using a similar method should not be possible for GAN and WGAN based architectures as well. One reason why AE based architectures 
might allow better training of post processing steps is however 
the higher correlation between real input and fake samples via
the latent space embedding. Nonetheless, the ability of the BIB-AE framework to make use of this post processing setup 
motivates future studies of this rather novel architecture for calorimeter shower generation.

\subsection{Computational Performance}

Beyond  the physics performance of our generative models, discussed in the previous section, the major argument
for these approaches is of course the potential gain in production time. To this end, we benchmark the per-shower generation time both on CPU and GPU hardware architectures. In Table~\ref{table:bench}, we provide the performance for 4 (3) batch sizes for the WGAN\footnote{The time evaluation of the GAN network is not reported since the generator architecture is very similar to the WGAN.} (BIB-AE).
We observe a speed-up by evaluating generative models on GPU vs. Geant4 on CPU of up to almost a factor of three thousand. 
Moreover, the evaluation time of our generative models is independent of the incident photon energy while this is not the case for the Geant4 simulation.


\begin{table*}
\centering
\label{table:bench}
\caption{Overview of computational performance of WGAN and BIB-AE model, compared to Geant4 full simulation.  Evaluated on both a single core of a  Intel\textsuperscript{\tiny\textregistered} Xeon\textsuperscript{\tiny\textregistered} CPU E5-2640 v4 (CPU) and NVIDIA\textsuperscript{\tiny\textregistered} V100 with 32~GB of memory (GPU). Numerical values
represent the mean and standard deviation of 25 runs.}
 \begin{tabular}{||c c l | l c| l c||} 
 \hline
 Simulator & Hardware & Batch Size& 15 GeV & Speed-up & 10-100 GeV Flat & Speed-up \\ [0.5ex] 
 \hline
Geant4     & CPU      & N/A       &  $1445.05\pm{19.34}$ ms          & -                   & $4081.53\pm{169.92}$ ms & - \\
\hline
WGAN & CPU       & 1            &  $64.34\pm{0.58}$ ms   & \textbf{x23}           &   $63.14\pm{0.34}$ ms & \textbf{x65}\\
 &        & 10           &  $59.53\pm{0.45}$ ms  & \textbf{x24}            &   $56.65\pm{0.33}$ ms  &\textbf{x72}\\
 &        & 100          &  $58.31\pm{0.93}$ ms  & \textbf{x25}           &  $58.11\pm{0.13}$ ms &\textbf{x70}\\
 &        & 1000         &  $57.99\pm{0.97}$ ms  & \textbf{x25}           &  $57.99\pm{0.18}$ ms &  \textbf{x70}\\
\hline
BIB-AE & CPU  & 1              & $426.60\pm{3.27}$ ms & \textbf{x3}      &  $426.32\pm{3.62}$ ms & \textbf{x10} \\
 &   & 10              & $422.60\pm{0.26}$ ms & \textbf{x3}      &  $424.71\pm{3.53}$ ms & \textbf{x10}\\
 &   & 100              & $419.64\pm{0.07}$ ms & \textbf{x3}      &  $418.04\pm{0.20}$ ms  & \textbf{x10} \\
\hline
WGAN & GPU       & 1              &  $3.24\pm{0.01}$ ms     &  \textbf{x446}   & $3.25\pm{0.01}$ ms  &  \textbf{x1256}      \\
 &      & 10             &  $6.13\pm{0.02}$ ms     &  \textbf{x236}   & $6.13\pm{0.02}$ ms  & \textbf{x666}        \\
 &       & 100            &  $5.43\pm{0.01}$ ms     &  \textbf{x266}   & $5.43\pm{0.01}$ ms  &  \textbf{x752}         \\
 &        & 1000           &  $5.43\pm{0.01}$ ms     &  \textbf{x266}   & $5.43\pm{0.01}$ ms  &  \textbf{x752}       \\
\hline
BIB-AE & GPU   & 1              & $3.14\pm{0.01}$ ms & \textbf{x460}      &  $3.19\pm{0.01}$ ms & \textbf{x1279} \\
 &   & 10              & $1.56\pm{0.01}$ ms & \textbf{x926}      &  $1.57\pm{0.01}$ ms & \textbf{x2600}\\
 &    & 100              & $1.42\pm{0.01}$ ms & \textbf{x1017}      &  $1.42\pm{0.01}$ ms  & \textbf{x2874}\\

\hline
\end{tabular}
\end{table*}


\section{Conclusion}

\label{closing}

The accelerated simulation of calorimeters with generative 
deep neural networks is an active area of research. 
Early works~\cite{CaloGAN2,CaloGAN,CaloGAN3} 
established generative networks as a fast and 
very promising tool for particle physics and 
simulated the positron, photon, and charged pion response of an idealised perfect calorimeter with 3 layers
and a total of 504 cells ($3 \times 96$, $12\times 12$, and $12 \times 6$).

Using the WGAN architecture and an energy constrainer network~\cite{ErdmannWGAN2} allowed the correct simulation
of the observed total energy of electrons for a calorimeter consisting of seven layers with a total of 1,260 cells ($12\times 15$ cells per layers). However, a mismodelling of individual cell energies below 10 MIPs, also leading to an observed deviation in the hit multiplicity distribution, was observed and studied. Our implementation of a WGAN based on~\cite{ErdmannWGAN2} reproduces this effect (see Fig.~\ref{fig:results_diffdist1}~(left)).
The proposed BIB-AE architecture with additional MMD loss term and Post Processor Network leads to a reliable description
of low energy deposits.

The ATLAS collaboration also reported the accurate simulation of high-level observables for photons in a four-layer
calorimeter segment with a total of 276 cells ($7 \times 3$, $57 \times 4$, $7\times 7$ and $7 \times 5$) using a VAE architecture~\cite{ATLAS_Gen2}
and 266 cells using a WGAN~\cite{ATLAS_Gen3}.
Recent progress was made applying a GAN architecture to simulating electrons in a high granularity calorimeter prototype~\cite{Sofia}. The considered detector consists of 25~layers with $51\times51$ cells
per layer, leading to a total of 65k cells to be simulated. On this very challenging problem, good agreement with Geant4 was achieved for a number of differential distributions and correlations of high-level observables. Specifically, the per-cell energy distribution was not reported, however the disagreement in the hit multiplicity again implies a mismodeling of the MIP peak region.

Our specific contribution is the first high fidelity simulation
for a number of challenging quantities relevant for 
downstream analysis, including  the overall energy response and per-cell energy distribution around the MIP peak,
for a realistic high-granularity calorimeter. This is made possible by the first application of the BIB-AE architecture 
--- unifying GAN and VAE approaches --- in physics. Modifications to this architecture, specifically an
additional kernel-based MMD loss term and a Post Processor Network, were developed.
These improvements can potentially also be applied to other generative architectures and models.
Planned future work includes the extension of this approach to also cover
multiple particle types, incident positions and angles towards a complete, fast,
and physically reliable synthetic calorimeter simulation.

\begin{acknowledgements}
The authors would like to thank
Martin Erdmann, Tobias Golling, Tilman Plehn, David Shih,
and Slava Voloshynovskiy for encouraging discussions and for providing valuable feedback on the manuscript.
We especially thank Ben Nachmann for his suggestions to improve the GAN training.
We would also like to thank the Maxwell and National Analysis Facility (NAF) computing
centers at DESY for the smooth operation and technical support.
E. Buhmann is funded by the German Federal Ministry of Science and Research (BMBF) via 
\textit{Verbundprojekts 05H2018 - R\&D COMPUTING
(Pilot\-maß\-nah\-me ErUM-Data) Innovative Digitale
Technologien f\"ur die Erforschung von Universum und
Materie}.
S. Diefenbacher is funded by the Deutsche Forschungsgemeinschaft (DFG, German Re\-search Foundation) 
under Germany’s Excellence Strategy – EXC 2121  ``Quantum Universe" – 390833306. 
E. Eren is funded through the Helmholtz Innovation Pool project AMALEA that provided a stimulating scientific environment for parts of the research done here.

\end{acknowledgements}

\section*{Conflict of interest}
The authors declare that they have no conflict of interest.

\newpage

\begin{appendix}

\section{Network architectures and training procedure}

The network architectures of generative models have a large number of moving parts 
and the contributions from various generators, discriminators, and critics need to
be carefully orchestrated to achieve good results. In the following we provide
details of the implementation and training for the GAN, WGAN, and BIB-AE models.
Due to the high computational cost of the studies --- e.g. the BIB-AE was trained for a
total of four days in parallel on four NVIDIA Tesla V100 (32 GB) GPUs --- 
no systematic tuning of hyperparameters was performed. For all architectures
a good modelling of the Geant4 training distributions was used as stopping criterion.
All architectures are implemented in PyTorch~\cite{pytorch} version 1.3.

\label{hyper}

\subsection{GAN Training}

Our implementation of the simple GAN is inspired by~\cite{CaloGAN,CaloGAN2,CaloGAN3} and it should serve
as an easy to implement baseline model consisting of a generator and a discriminator. 
In total, the generator has 1.5M trainable weights and the discriminator has 2.0M weights.
We therefore did
not consider additional modifications to the GAN approach
such as training with a gradient penalty term. 

The generator network of the GAN consists of 3-di\-men\-sional transposed convolution layers with batch normalization. It takes a noise vector of length 100, uniformly distributed from -1 to 1,  and the true energy labels $E$ as inputs. A first transposed convolution with a $4^3$ kernel (stride 1) 
is applied to the noise vector multiplied by $E$.
The main transposed convolution consists of four layers. The first three layers have a 
kernel size of $4^3$ (stride 2) 
followed by batch normalization. The final layer has a kernel size of $3^3$ (stride 1).
All layers use ReLU~\cite{ReLU} as activation function.

The discriminator uses five 3-dimensional convolution layers followed by two fully connected layers with 257 and 128 nodes respectively. The convolution layers use a  $3^3$ kernel.
The stride is 2 for all convolutional layers.
Batch normalisation~\cite{batchnorm} is applied after each convolution except in the first and last layer. 
We flatten the output of the convolutions and concatenate it the with input energy
before passing it to the fully connected layers. Each fully connected layer except the  final one uses LeakyReLU~\cite{LReLU} (slope: $-0.2$) as an activation function. The activation in the final layer is sigmoid.

For training, we use the Adam optimizer~\cite{adam} (learning rate $2 \cdot 10^{-5}$). The training process starts from updating the discriminator for real and fake showers. After that we freeze the parameters of the discriminator and update the generator with a new generated batch of fake showers. The generator and discriminator are trained alternating until the training is stopped after 125k weight updates --- corresponding to approximately 6 epochs --- when good modelling of the control distributions is achieved.


\subsection{WGAN Training}

The WGAN architecture, based on~\cite{ErdmannWGAN,ErdmannWGAN2}, consists of 3 networks: one generator with 3.7M weights, 
one critic with 250k weights, and one constrainer network with 220k weights.
The critic network starts with four 3D convolution layers with kernel sizes ($X$,2,2) with $X=10,6,4,4$ which have 32, 64, 128, and 1 filters respectively. LayerNorm~\cite{layernorm} layers are sandwiched between the convolutions. 
After the last convolution, the output is concatenated with the $E$ vector required for $E-$conditioning. After that, it is flattened and fed into a fully connected network with 91, 100, 200, 100, 75, 1 nodes. Throughout the critic, LeakyReLU (slope: $-0.2$) is used as activation function.

The generator network takes a latent vector $z$ (normally distributed with length 100)
and true $E$ labels as input and separately passes them through a 3D transposed convolution layer using a $4^3$ kernel with 128 filters. After that, the outputs are concatenated and processed through a series of four 3D transposed convolution layers (kernel size $4^3$ with filters of 256, 128, 64, 32).  LayerNorm layers along with ReLU activation functions are used throughout the generator.

The energy-constrainer network is similar to the critic: three 3D convolutions with kernel sizes $3^3$, $3^3$ and $2^3$ along with 16, 32, and 16 filters are used. The output is then fed into a fully connected network with 2000, 100, and 1 nodes. LayerNorm layers and LeakyReLU (slope: -0.2) are sandwiched in between convolutional layers.

The WGAN is trained for a total of 131k weight updates which corresponds to 20 epochs. The
generator and critic network are trained using the Adam optimizer with an initial
learning rate of $10^{-4}$. The learning rate is decreased by a factor of 10 each after
the first 50k and after a total of 100k iterations. For the critic, the initial learning
rate is $10^{-5}$. It is reduced by a factor of 10 after 50k iterations. 
Finally, the constrainer network is trained using stochastic gradient descent~\cite{SGD} with a learning rate 
of $10^{-5}$. After 30k iterations, the constrainer weights are frozen.
The training of the WGAN took one week on three NVIDIA Tesla V100 GPUs.

\subsection{BIB-AE Training}

\begin{figure*}[h]
    \centering
    \includegraphics[width=0.23\textwidth]{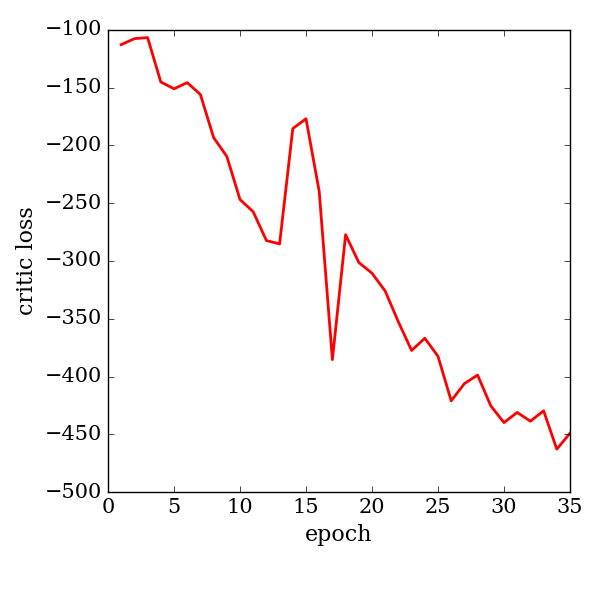}
    \includegraphics[width=0.23\textwidth]{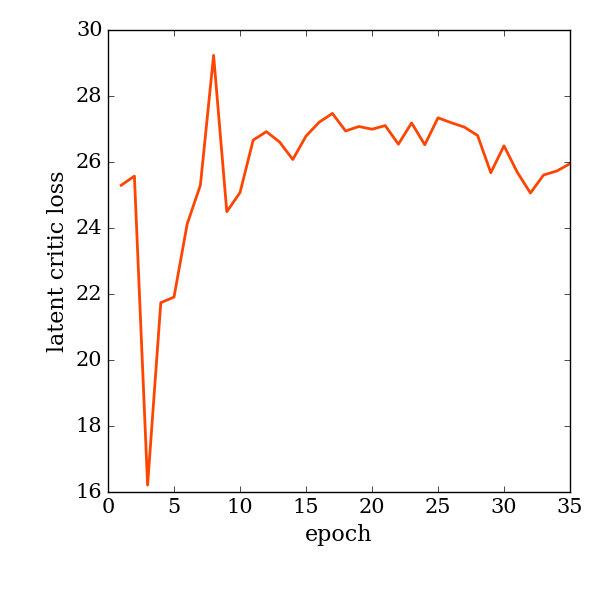}
    \includegraphics[width=0.23\textwidth]{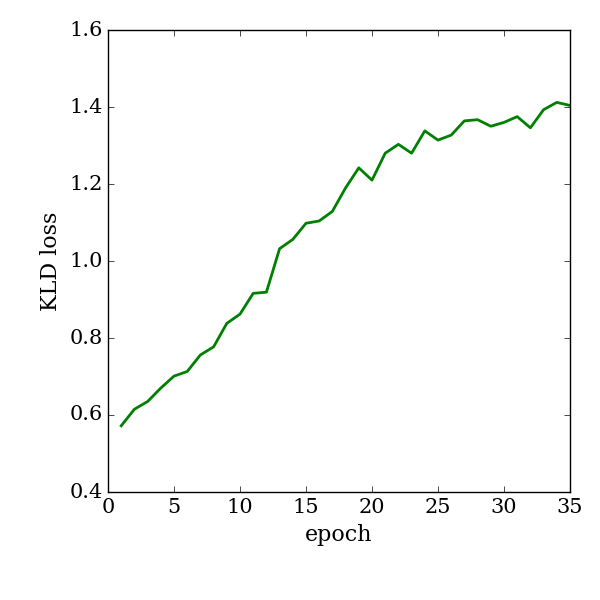}
    \includegraphics[width=0.23\textwidth]{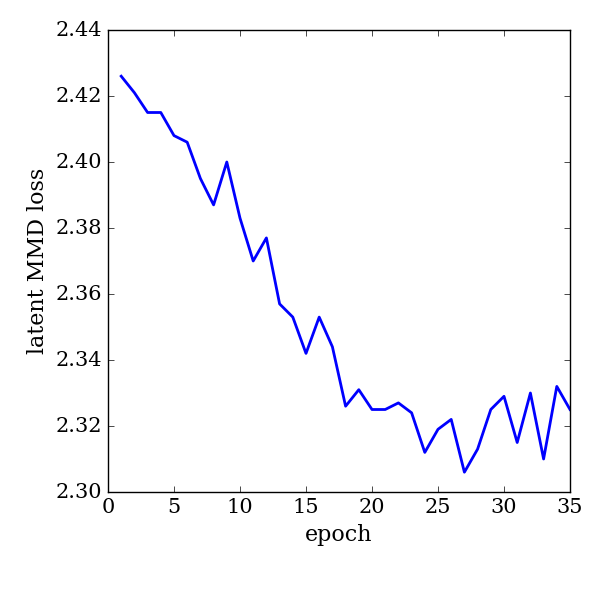}
    \caption{
    Evolution of the indiviual loss contributions during the BIB-AE training. From left to right: critic loss, latent critic loss, KLD loss and latent MMD loss.}
    \label{fig:BIBAE_loss}
\end{figure*}

Our implementation of the BIB-AE architecture consists of an encoder and a decoder, a latent space critic, 
a pair of critic and difference critic, and a network for post processing, and has 71M weights in total.
Of these, 35M weights are used by the encoder. This is a significantly larger number of weights than what can be found in the GAN and WGAN models, however this can largely be attributed to the use of fully connected layers in the BIB-AE, while both GANs are almost purely convolutions. Regardless of this weight discrepancy both models remain comparable, since their total computing time is in the same order of magnitude, as can be seen in Table \ref{table:bench}.

The encoder consists of four 3-dimensional convolution layers with kernel size $4^3$, $4^3$, $4^3$ and $3^3$, stride 2, 2, 2 and 1 and 8, 16, 32 and 64 filters. After each convolution LayerNorm is applied. The final convolution has an output shape of $64 \times 5 \times 5 \times 5$. This output is flattened, concatenated with the true energy label, and passed to a series of dense layers with 8001, 4000, 32 and $2 \times 24$ nodes. The two sets of 24 final outputs are interpreted as $\mu$ and $\sigma$ and are used to define 24 Gaussian distributions. We sample once from each Gaussian to form the latent representation of the input shower. These 24 values are passed to the decoder. 

The decoder takes the 24 latent-samples and concatenates them with 488 points of random Gaussian noise as well as the true energy label. The resulting tensor is then passed to dense layers with 513, 768, 4000 and 8000 nodes. We reshape the output of the dense layers to $8 \times 10 \times 10 \times 10$. Using two transposed convolution layers with kernel sizes $3^3$ and $3^3$, strides 3 and 2, and 8 and 16 filters respectively this is upsampled to $16 \times 60 \times 60 \times 60$ and then reduced back down to $8 \times 30 \times 30 \times 30$ by a kernel-size $2^3$, stride 2 convolution. This is followed by four more convolutions, all with kernel-size $3^3$ and stride 1 with 8, 16, 32, and 1 filters respectively. Once again each (transposed) convolution except for the last one is followed by LayerNorm. Both encoder and decoder use LeakyReLU as intermediate activation functions. The final encoder layer has a linear, the final decoder layer a ReLU activation. 


The BIB-AE latent space critic is a fully connected network with 1, 50, 100, 50, and 1 nodes using LeakyReLU activation. The critic is trained using samples from a Normal distribution as true data and using the latent space samples as fakes. Each of the 24 sampled latent space variables is passed individually to the critic.

The BIB-AE critic and difference critic are built as a combined network with four input streams. The first stream takes the $30 \times 30 \times 30$ shower image as input and applies 3 convolutions with kernel-size $3^3$, $3^3$, and $3^3$, stride 2, 2, and 1, and 128, 128, and 128 filters, reducing the input to $128 \times 4 \times 4 \times 4$. The convolutions are interspersed with LayerNorms. The convolutional output is flattened and passed to a dense layer with 64 output nodes. The second stream is nearly identical to the first one, except the input is scaled 
by adding one and applying the natural logarithm. The third stream consists of a single dense layer with $30^3=27,000$ input and 64 output nodes. The input to this stream is the flattened difference between the reconstructed image and the original image. 
Finally, we use the true energy label as input to the fourth stream. It consists of one dense layer with one input and 64 outputs. 

The 64 outputs from each of the four streams are concatenated and passed to a final set of dense layers with 256, 128, 128, 128, 1 nodes. We once again use LeakyReLU everywhere except for the final layer, which has a linear activation. During training the first two streams receive Geant4 images as real data and reconstructed images as fakes. The third stream receives Geant4-Geant4 as real and Geant4-reconstructed as fake. The fourth stream always receives the true energy label.

The Post Processor Network also has two streams. The first takes a $30 \times 30 \times 30$ image as its input and applies a kernel-size $1^3$, stride 1 convolution with 128 filters. The second one takes the true energy label and the sum over all pixels in the input image as its input.  These are passed to dense layers with 2, 64, 64, 64 nodes, the output of which is expanded to a $64 \times 30 \times 30 \times 30$ shape. The tensor is then concatenated along the filter dimension with the $128 \times 30 \times 30 \times 30$ output of the first stream. The combined object is passed to five more convolutions, all with kernel-size $1^3$, stride 1 and 128, 128, 128, 128,  and 1 filters. As before, convolutions are interspersed with LayerNorms. We use LeakyReLU save for the last layer which uses a linear activation. The use of kernel-size $1^3$ means that the same function is applied to every pixel value. However the intermittent LayerNorms cause the precise functions to be different for each individual shower as well as for each pixel within the showers. As a result, each shower has its own set of 27000 functions that behave very similarly, but are still tailored to each of the 27000 possible pixel positions.

The setup is initially trained for 35 epochs without the Post Processor, the evolution of the individual loss contributions during this training is shown in Fig.\ref{fig:BIBAE_loss}. The initial learning rates are $0.5\times10^{-3}$ for encoder, decoder and the critic, and $2.0\times10^{-3}$ for the latent critic. All learning rates decay by $0.95$ after each epoch. For each encoder/decoder update we update the critics 5 times. After these 35 epochs we train the Post Processor for one epoch using only the MSE term. This ensured the Post Processors baseline behaviour is to make as little changes to the images as possible. For three subsequent epochs the Post Processor is trained using a combination of MSE and MMD, with the same learning rate as the encoder/decoder. The initial 35 epochs of training took 3 days on four NVIDIA Tesla V100 (32 GB) GPUs and the Post Processor training lasted for one additional day.
We save checkpoints after each epoch. A composite figure of merit combining a number of 1D distributions was used to evaluate when stopping was warranted and to select which checkpoint shows the best agreement with the training data.

\end{appendix}

\bibliographystyle{bib_style}
\bibliography{4_BIB.bib}   

%
%

\end{document}